# The Underlying Mechanisms of Time Dilation and Doppler Effect in Curved Space-Time

Benliang Li,[1,*] Hailiang Zhang,[2] and Perry Ping Shum,[2]
[1]*School of physical Science and Technology, Southwest Jiaotong University, Chengdu, 610031, China*
[2]*COFT, School of EEE, Nanyang Technological University, 50 Nanyang Avenue, Singapore 639798*
*Email: libenliang732@swjtu.edu.cn

In this paper, we theoretically investigate the time dilation and Doppler effect in curved space-time from the perspective of quantum field theory (QFT). A Coordinate Transformation which Maintains the Period of Clocks is introduced, and such coordinate transformation is named as CTMPC throughout this paper. By analogy with the Lorentz transformation in Minkowski space-time, CTMPC is a correct transformation in curved space-times in a sense that it shows the correct relation between the time measured by the two observers, moreover, Lorentz transformation is just a special case of CTMPC applied in Minkowski space-time. We demonstrate that the Coordinate Transformation which Maintains the Local Metric (CTMLM) is one CTMPC, while the mathematical forms of physics formulas in QFT will be maintained. As applications of CTMLM, the time dilation and Doppler effect with an arbitrary time-dependent relative velocity in curved space-time are analysed. For Minkowski space-time, the time dilation and Doppler effect agree with the clock hypothesis. For curved space-time, we show that even if the emitted wave has a narrow frequency range, the Doppler effect may, in general, broaden the frequency spectrum and, at the meantime, shift the frequencies values. These new findings will deepen our understanding on the nature of space-time and the Doppler effect in curved space-time, they may also provide theoretical guidance in future astronomical observations.

## I. Introduction

In Minkowski space-time, the Lorentz transformation is the valid coordinate transformation between two coordinate systems set by two observers respectively. In curved space-time, physicists usually define new coordinate parameters in order to make the mathematical calculations become less complicated, and the newly introduced coordinate parameters may not be able to represent the physical space and time. Take Eddington–Finkelstein coordinates and tortoise coordinate for example, the new defined parameter $v = t + r^*$ may not be able to represent the time measured by clocks. Therefore, similarly as the Lorentz transformation in Minkowski space-time, for two observers, who are carrying two identical clocks respectively, moving in curved space-times with arbitrary velocities, one way wonder whether we can find the correct coordinate transformation which agrees with the time measured by the two clocks? To answer this question, in this paper, we introduce a new Coordinate Transformation that Maintains the Period of Clocks (CTMPC) then apply this coordinate transformation to deal with the time dilation and Doppler effect in curved space-times.

The quantum field theory (QFT) in curved space-time is to investigate the behaviour of quantum particles with fixed space-time metric. Since some operators in QFT (such as Dirac $\gamma^{\mu}$ factor in QED and covariant derivative $\nabla_{\mu}$) are metric dependent, as a result, the mathematical expressions of waves and the formulas describing the interactions between particles will carry the metric dependence. In our previous work [1], we argued that the formation of atomic structures can be theoretically deduced from the QFT. Therefore, the atomic spectra will carry the information about the local metric, and the gaps between the energy levels of the atom will vary with the position of the atom in curved space-time. In order to analyze the energy levels of the atoms located at different positions in curved space-time, one can study the local curvature inside of the atoms to obtain the shifted energy spectrum, which has been a research topic under investigation using perturbation theory [2–7].

In modern cosmology, the electromagnetic signals coming from moving astronomical objects deliver the exhaustive physical information about the intriguing phenomena going on in the universe. Providing a solid theoretical instruction to analyze these signals is crucial in leading to new and important discoveries [8–10]. As a diagnostic tool, the Doppler shift measurement is one of the most important spectroscopic measurements made in astronomy. Achieving a complete understanding of the Doppler effect in the curved space-time becomes essential in studying the numerous astrophysical phenomena and some progress has been made by many researchers in the past [11–18]. For the Doppler effect in curved space-time, one needs to solve the wave equations of QFT to examine the evolution of waves during the propagations. Besides, since atoms usually are used as emitters or receivers of waves, the interactions between the propagating waves with the atoms whose energy levels might be shifted due to the motion in curved space-time need to be well understood. Because of these reasons, the combination of general relativity with QFT to investigate the Doppler effect becomes a necessity, and some new features that have not been considered in previous works may arise from it.

In this paper, we investigate the influence of space-time curvature on two atoms located far away from each other such that the metric value inside the atoms can be taken as constants. In order to analyse the behaviour of these atoms, we introduce a new coordinate system $(t', x')$ which maintains the local metric the same as that in the original coordinate $(t, x)$, and such Coordinate Transformation which Maintains the Local Metric is abbreviated as CTMLM throughout this paper. As a result of CTMLM, the quantum states of the atoms in two different coordinate systems take the same mathematical forms, therefore, CTMLM can be proved to be one type of CTMPC. Meanwhile, as applications of CTMLM, the time dilation and Doppler effect with an arbitrary relative velocity in curved space-time will be studied. This paper is organized as follows.

This paper is organized as follows. In section II, first we introduce CTMPC and show CTMPC is a correct coordinate transformation in curved space-times in analogy to the fact that Lorentz transformation is a correct transformation in Minkowski space-time. Then we demonstrate that CTMLM is one type of CTMPC. Meanwhile, the significance of CTMPC and CTMLM is discussed in this section. In sections III, IV and V, we apply CTMLM to study the time dilation and Doppler effect in Minkowski, Schwarzschild and FRW space-times, respectively. The results show that the time dilation in these space-times agree with the clock hypothesis while the Doppler effect in curved space-times displays some new features which have not been investigated by researchers before. For strong gravitational fields such as in the vicinity of black holes, the Doppler effect can shift the value of wave frequency and, at the same time, broaden the spectrum of the emitted signals. Under certain conditions and approximations, the Doppler-shifted frequencies are given in Schwarzschild and FRW space-times. These findings can provide a test for QFT in curved space-time and meanwhile may lead to important new modern astrophysical discoveries. In this paper, natural units in which $\hbar = G = c = 1$ will be used.

## II. CTMPC and CTMLM

In physics, Coordinate system is a mathematical construct to measure the position and the time of a physical event. Physicists construct mathematical equations and formulas to describe the physics laws, and one physics law may take different mathematical forms corresponding to different coordinate systems that were applied. For example, in coordinate $(t, x)$, the motion of a rigid body with mass m attached with a spring with spring constant $k$ can be given by a mathematical equation as

$$k(x - x_0) = \mathrm{m}\frac{d^2 x}{dt^2} \tag{2.1}$$

A new coordinate system $(t', x')$ with unit length and unit time rescaled can be adopted as

$$\begin{aligned} t' &= 2t \\ x' &= 4x \end{aligned} \tag{2.2}$$

Therefore, the mathematical form of Eq. (2.1) written in coordinate system $(t', x')$ becomes

$$\frac{1}{4}k(x' - x'_0) = \mathrm{m}\frac{d^2 x'}{dt'^2} \tag{2.3}$$

which is slightly different from Eq. (2.1) by a constant factor 1/4 in front of the equation despite the fact that these two equations represent the same physics law. In Euclidean physics, such confusion can be avoided by the unification of basic units, such as length unit and time unit in coordinate systems, that were applied to describe the physics laws. However, things may become tricky in general relativity theory where the time parameter and space parameter can be mixed, in other words, we may adopt another coordinate system $(t'', x'')$ as

$$\begin{aligned} t'' &= t + x \\ x'' &= x \end{aligned} \tag{2.4}$$

Eq. (2.1) written in coordinate system $(t'',x'')$ becomes

$$k(x''-x''_0)(1-\frac{dx''}{dt''})^3 = \mathrm{m}\frac{d^2x''}{dt''^2} \tag{2.5}$$

which takes an entirely different mathematical form compared with Eq. (2.1). One may wonder what are the physical meanings of mathematical parameters $(t'',x'')$ as defined in Eq. (2.4), and under what conditions they indeed represent the physical time and space? Keeping this question in mind, let us move on first then come back to this question later.

Clocks are mechanical or electronic devices that keep the track of time by counting the number of the periodic phenomena occurred. In modern atomic clocks, the atom Caesium133 is used and the transition frequency $\omega$ is defined as $\omega = 2\pi N$ Hertz where $N = 9192631770$, or equivalently to say that this atomic clock will display 1 second after $N$ cycles of the radiation counted inside of this clock. This indicates that the value of the time parameter $t$ in coordinate system $(t,x)$ is fixed, since it must agree with the time displayed by the clock. It also indicates that the coordinate transformation between two observers cannot be arbitrary, and we already know that the Lorentz transformation is the only correct transformation in Minkowski space-time. For general relativity, it states that there is not one reference point (or perspective) in the universe that is more favored than another. This might be a tricky statement since one needs to be careful when two coordinate systems are applied to describe our universe. The reason is as follows.

First, let us look at one example. In Schwarzschild space-time, the Eddington–Finkelstein coordinate are founded upon the tortoise coordinate in which $r^*$ is defined as

$$r^* = r + 2GM\ln\left|\frac{r}{2GM}-1\right| \tag{2.6}$$

And the ingoing Eddington–Finkelstein coordinates are obtained by replacing the coordinate $t$ with the new coordinate

$$t^* = t + r^* \tag{2.7}$$

as the new "time" parameter in Eddington–Finkelstein coordinate system. In Schwarzschild space-time, suppose Alice who carries atomic clock A sets the coordinate $(t,r)$ to describe the time displayed by clock A. Bob, who is located somewhere in Schwarzschild space-time and moving with time-dependent velocity $v_B(t)$ measured by Alice, carries an identical atomic clock B. Now we come back to the question asked earlier: under what conditions the parameter $t^*$ in Eq. (2.7) indeed stands for the time displayed by clock B? In other words, under what conditions the coordinate transformation Eq. (2.6) and Eq. (2.7) is the correct transformation between the two observers i.e., Alice and Bob? The answer partly rely on the spatial locations of Alice $r_A$ and Bob $r_B$ in Schwarschild space-time and the relative velocity $v_B(t)$. That is, one may find two special locations of Alice $r_A$ and Bob $r_B$, and one special velocity $v_B(t)$ such that the coordinate transformation between Alice and Bob indeed satisfy Eq. (2.6) and Eq. (2.7); to be more specific, the relationship between the time $t_A$ displayed by clock A and the time $t_B$ displayed by clock B satisfies Eq. (2.7). However, it is also possible that the time displayed by the two clocks $t_A$ and $t_B$ never satisfy Eq. (2.7) even by exhausting all possible values of $r_A$, $r_B$ and $v_B(t)$, under such situation, we can conclude that the transformation Eq. (2.6) and Eq. (2.7) is a mathematically technique that carries no physical meaning, since we have no idea what physical meaning for symbol $t^*$ represents, the same conclusion applied to Eq. (2.4). At this stage, one may wonder how to find the correct coordinate transformation between Alice and Bob with arbitrary values of $r_A$, $r_B$ and $v_B(t)$ in curved space-times such that the time transformation shows the correct relationship between the time displayed by clock A and clock B? In order to answer this question, first let us look at one simple example.

Suppose Alice and Bob are carrying two identical pendulum-clocks which are named as clock A and clock B respectively, both of them are statically located on earth. The period of the pendulum-clock is given as

$$T = 2\pi\sqrt{\frac{L}{g}} \tag{2.8}$$

where $L$ is the length of the pendulum and $g$ is the local acceleration of gravity on earth. Alice who is carrying clock A sets a coordinate $(t,x)$, Bob who is carrying clock B is located at $x_B$, note that $x_B$ is measured in Alice's coordinate $(t,x)$. Alice and Bob are situated at two places with equal local acceleration of gravity, i.e., $g_A = g_B$. Suppose Alice and Bob never met before, and they both do not know the SI unit system or other unit systems in modern world, to be more specific, they did not share the common units in their coordinate systems. In other words, this means that for the two identical objects A and B carried by Alice and Bob respectively, Alice may measure the length of object A as one meter while Bob measures the length of object B as four meters which is similar as what is shown in Eq. (2.2), since they did not make any agreement over the unit of length applied in their coordinate systems. In Alice's coordinate $(t,x)$, the length $L$ of clock A takes the value $L_A = 10m$ and local acceleration $g_A$ takes the value $g_A = 10m/s^2$, therefore, $T = 6.28$ and clock A will display 6.28 seconds after one period of pendulum is completed. Bob, who is carrying clock B which is identical with clock A, does not know what value of length $L$ is taking in Alice's coordinate $(t,x)$. If he sets a coordinate system $(t',x')$ in which $L'_B = 1m$ and $g'_B = 10m/s^2$, clock B will display $\frac{6.28}{\sqrt{10}}$ seconds after one period of pendulum is completed. After $N_A$ periods counted by clock A, it will display the time measured as 6.28 $N_A$; clock B, at the same time, counts the same number of the periods completed as $N_B = N_A$ since the two clocks are identical and the local accelerations of gravity are the same. Therefore, the time displayed by clock B is $\frac{6.28N_B}{\sqrt{10}}$ which is $\sqrt{10}$ times smaller than the time displayed by clock A. Such difference of the time measured by the two clocks is due to the wrong coordinate system that Bob applies, not because of any physical laws such as gravity or high speed motions. In order to avoid such inconsistency caused by the wrong coordinate system that Bob applies, Alice and Bob need to adjust their clocks before the measurement. In practice, this is what people usually do, they compare and adjust two clocks located at one place before the experiment, then one clock is taken away from the other to conduct the experiment. People adjust the two identical clocks to make sure that they both display the same time value after one period is completed. For example, Alice can make a phone call to Bob and tell him that clock B needs to display the same time value $t' = 6.28$ seconds after one period is counted, therefore, Bob immediately realize that he must set a coordinate system $(t', x')$ with adjusted unit of length such that the length of the pendulum clock B $L_B$ takes the value $L'_B = 10m$, note that this value is measured in coordinate $(t', x')$. Therefore, the period of clock A $T_A$ measured in coordinate $(t, x)$ is equal to the period of clock B $T'_B$ measured in coordinate $(t', x')$, i.e., $T_A = T'_B = 6.28$. In this way, after a while, the difference between the values of the coordinate parameters ( $t$ and $t'$ ) becomes the difference between the two numbers of periods completed by two clocks, that is,

$$\frac{t'}{t} = \frac{T'_B N_B}{T_A N_A} = \frac{6.28 N_B}{6.28 N_A} = \frac{N_B}{N_A}. \tag{2.9}$$

Note that if Alice and Bob meet and compare their two clocks, what actually being compared are the two numbers ( $N_A$ and $N_B$ ) of the periods counted by clock A and clock B, the difference between the two numbers of the periods reflects the difference between the time measured by the two clocks, and such time difference is caused by the physical laws (gravity or the fast speed motion), not because of the wrong coordinate system that Bob applies. At this stage, we can make a conclusion: If Alice and Bob set their coordinate systems $(t, x)$ and $(t', x')$ such that the period of clock A $T_A$ which is measured in coordinate $(t, x)$ takes the same value as the period of clock B $T'_B$ which is measured in coordinate $(t', x')$, i.e., $T_A = T'_B$, then the coordinate transformation between $(t, x)$ and $(t', x')$ shows the correct relationship between the time displayed by clock A and clock B, as indicated by Eq. (2.9). For convenience only, we name this Coordinate Transformation which Maintains the Period of Clocks as CTMPC throughout this article.

CTMPC is drawn from the analysis of pendulum clocks, we can ask ourselves whether it applies to atomic clocks moving in curved space-times? In order to answer this question, we notice that for the atomic clock, the transition frequency $\omega$ is defined as $\omega = 2\pi N$ Hertz where $N$ is a fixed number for all identical atomic clocks. Thus, the period of the atomic clock can be given as $T = \frac{2\pi}{\omega} = \frac{1}{N}$ seconds, which means that the clock will display 1 second after $N$ cycles of the radiation counted inside of this clock regardless where and when this clock is located, or how fast this clock is moving. As a result, whoever carries this clock must set a coordinate system $(t,x)$ in which the period of the clock takes the value as $T=\frac{1}{N}$. Therefore, as can be concluded, CTMPC holds for any clocks moving in any space-times. For example, in curved space-time, suppose Alice and Bob are carrying two atomic clocks which name as clock A and clock B respectively. Alice sets a coordinate $(t,x)$ in which the value of time parameter $t$ is the same as the time displayed by clock A. Bob is moving with speed $v_B(t)$ which is measured in Alice's coordinate $(t,x)$. Bob sets another coordinate $(t',x')$ which satisfies CTMPC with Alice's coordinate $(t,x)$. After a while, $N_A$ cycles of the radiation is counted inside of clock A and clock A will display the time value $t=N_A T=\frac{N_A}{N}$; meanwhile, $N_B$ cycles of the radiation is counted inside of clock B, thus, in Bob's coordinate $(t',x')$ which satisfies CTMPC, the value of the time parameter $t'$ is given by $t'=N_B T=\frac{N_B}{N}$. Note that the time dilation between the two clocks can be given as

$$\frac{t'}{t}=\frac{N_B}{N_A} \tag{2.10}$$

since the common factor $N$ which is just the inverse of the period of the two clocks is canceled, as indicated by CTMPC. Therefore, Eq. (2.10) implies that the time difference between the two coordinate systems reflects the difference between the two numbers, i.e., the numbers of the periods completed by two clocks $N_A$ and $N_B$, and the difference between these two numbers is caused by the physics laws such as the inequality strength of local gravity where clock A and clock B are located and the relative velocity $v_B(t)$. As a result, Eq. (2.10) shows the correct relationship between the time measured by the two clocks, and CTMPC is the correct transformation that Alice and Bob need to follow when they set their coordinate systems.

In the following paragraphs of this section, we discuss a Coordinate Transformation that can Maintain the Local Metric (CTMLM). By keeping the local metric value, we show that the physical formulas of local QFT remain the same in the transformed coordinate, besides, we will demonstrate that CTMLM is one type of CTMPC.

For QFT in curved space-time $(t,x)$, we write the two-dimensional equation of motion for massless scalar field $\psi(t,x)$ as

$$g^{\mu\nu}(t,x)\nabla_\mu\nabla_\nu\psi(t,x)=0, \tag{2.11}$$

where $\nabla_\mu$ stands for the covariant derivatives and $g^{\mu\nu}(t,x)$ is the metric in $(t,x)$ coordinate system. Now if we use another set of coordinate $(t',x')$, we get the field equation as

$$g'^{\rho\sigma}(t',x')\nabla'_\rho\nabla'_\sigma\psi'(t',x')=0, \tag{2.12}$$

where

$$g'^{\rho\sigma}(t',x') = \frac{\partial \tilde{x}'^{\rho}}{\partial \tilde{x}^{\mu}} \frac{\partial \tilde{x}'^{\sigma}}{\partial \tilde{x}^{\nu}} g^{\mu\nu}(t,x) \tag{2.13}$$

and $\nabla'_{\rho} = \dfrac{\partial \tilde{x}^{\mu}}{\partial \tilde{x}'^{\rho}} \nabla_{\mu}$ with the defined symbol $\tilde{x}^0 \equiv t$ ( $\tilde{x}'^0 \equiv t'$ ) and $\tilde{x}^1 \equiv x$ ( $\tilde{x}'^1 \equiv x'$ ). Therefore, the operation $g'^{\rho\sigma}(t',x')\nabla'_{\rho}\nabla'_{\sigma}$ in coordinate $(t',x')$ will become equivalent with $g^{\mu\nu}(t,x)\nabla_{\mu}\nabla_{\nu}$ in coordinate $(t,x)$. Thus, we can see that the scalar field solution $\psi(t,x)$ in Eq. (2.11) is also a solution for Eq. (2.12) in coordinate $(t',x')$, that is,

$$g'^{\rho\sigma}(t',x')\nabla'_{\rho}\nabla'_{\sigma}\psi(t,x) = 0 . \tag{2.14}$$

Therefore, we have $\psi'(t',x') = \psi(t,x)$ (the value of the scalar function does not change). However, the mathematical structure of the solution $\psi'(t',x')$ may be different with $\psi(t,x)$, this is because $g'^{\rho\sigma}(t',x')$ may have a different mathematical structure compared with $g^{\mu\nu}(t,x)$. For the coordinate transformations that can maintain the mathematical structure (or mathematical form) of the metric $g^{\mu\nu}(t,x)$, the metric $g'^{\rho\sigma}(t',x')$ in coordinate $(t',x')$ remains unchanged such that $g'^{\rho\sigma}(T,X) = g^{\rho\sigma}(T,X)$ [ $g'^{\rho\sigma}(T,X)$ is the metric in $(t',x')$ coordinate with $t' = T, x' = X$ and $g^{\rho\sigma}(T,X)$ is the metric in $(t,x)$ coordinate with $t = T, x = X$ ] for any space-time value $(T,X)$. In this way, Eq. (2.14) becomes

$$g^{\rho\sigma}(t',x')\nabla'_{\rho}\nabla'_{\sigma}\psi(t,x) = 0 . \tag{2.15}$$

We also know from Eq. (2.12) that $\psi(t',x')$ is a solution of Eq. (2.15), i.e., we have

$$\psi(t',x') = \psi(t,x) . \tag{2.16}$$

Therefore, the mathematical form of the scalar function $\psi(t,x)$ remains unchanged. Thus, we can describe the same field using the same mathematical expression in different coordinate systems, the only difference is the value of the parameters. In fact, for the plane wave $e^{i(\omega t - kx)}$ as a scalar function which is a solution of Eq. (2.11) with Minkowski metric, the solution in $(t',x')$ coordinate that maintains the mathematical form of Minkowski metric becomes $e^{i(\omega' t' - k' x')}$ which has the same mathematical structure as $e^{i(\omega t - kx)}$, that is, the only difference between two solutions is the values of the four parameters $(t,x,\omega,k)$.

In practice, solving QFT equations involving curved metric $g^{\mu\nu}(t,x)$ extended in entire space-time would be highly difficult, especially when several fields are interacting such as QED and QCD in curved space-time. However, for local QFT phenomena (such as quantum states of atoms which only extends within nanometers and the interactions of atoms with background fields), we do not expect the space-time metric far way would have any influence on such QFT phenomena at this spot. Based on this observation, next we are going to discuss CTMLM and show that it is one CTMPC.

The atomic clock is a time-keeping device by counting the number of cycles of the radiated wave (for wave $e^{-i\omega t}$, the number of cycles completed is $n = \frac{\omega t}{2\pi}$ ) produced by transitions between the first excited state and the ground state of an atom. This atomic clock comprises two major parts: atoms and the radiation field produced by the transitions between the first excited state and the ground state, atoms and the radiation field are all confined within the clock device. Both the atomic transition and the evolution of the radiation field are physical phenomena that can be described by QFT. Therefore, from now on, we will treat the ticking of atomic clock as local field phenomena (local means within the size of the clock device) which can be described by QFT.

Now suppose there are two identical atomic clocks A and B, which are carried by Alice and Bob, respectively. Alice sets the coordinate $(t, x)$ to describe the physical phenomena which includes atomic transitions and evolutions of quantum fields inside of clock A, likewise, Bob sets the other coordinate $(t', x')$. Measured by Alice, Bob is located at $x_B(t)$ with a velocity $v_B(t) \equiv \frac{dx_B(t)}{dt}$. Given all the above conditions, now we can make a Motion-Hypothesis (MH):

**The motion of a clock (may be accelerated) does not exert influence on the QFT physical phenomena.**

In MH, the QFT physical phenomena include quantum states of atoms and the radiation fields generated by the transitions of atoms inside of the clock. Later on we will see that MH agrees with the clock hypothesis in special relativity, and MH is indeed a generalization of one of the postulates in special relativity: physics are the same in different inertial reference frames. And we just generalized this postulate to acceleration frames with a time-dependent relative velocity $v_B(t) \equiv \frac{dx_B(t)}{dt}$. However, such generalization does not include classical physics, we only refer to the microscopic physical phenomena arising from QFT. Moreover, MH is not only about the clocks, it is equivalent to state that if we confine an atom in some type of a box, the motion of the box does not affect the state of the atom and the interactions between the atoms with the background field. As a matter of fact, we would like to regard it as an approximation rather than a hypothesis since a super large acceleration of a box may even tear the atoms inside apart and this acceleration will clearly cause effects on the quantum states of atoms as well as the radiation field inside of the box. However, since the clock hypothesis has been verified with high precision by several experiments under a certain range of accelerations [19–23], as a convention, we still use the word "hypothesis" rather than "approximation". Next we are going to discuss the physical meaning of MH in details.

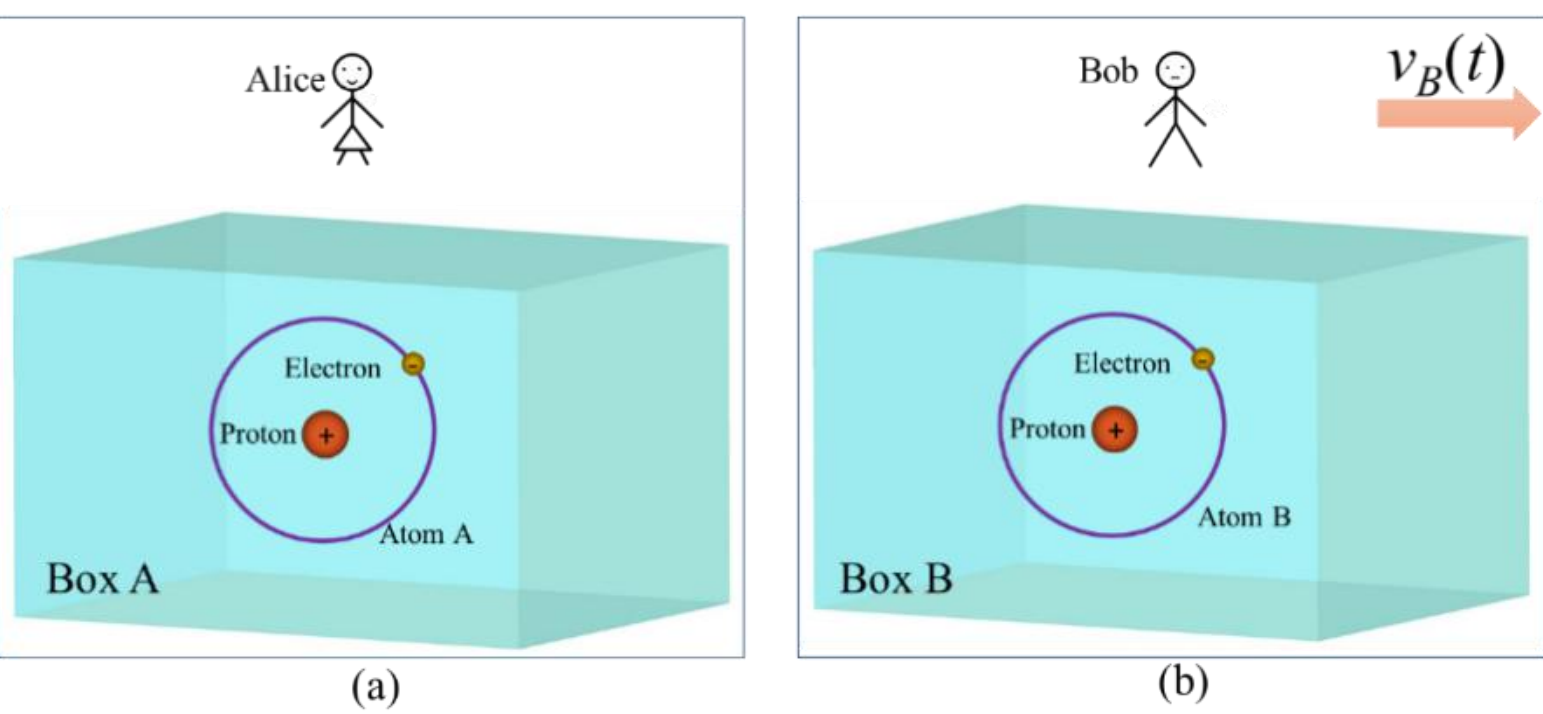

FIG. 1. Sketches of Box A and Box B, each one contains one hydrogen atom. (a) Box A is at rest. (b) Box B is moving with a velocity $v_B(t)$ measured in Alice's $(t,\vec{x})$ coordinate. If Bob sets another coordinate $(t',\vec{x}')$ that maintains the local metric where Box A occupies, then he can describe the quantum states of the atom B using the same formula that Alice applies to describe the states of the atom A.

As shown in Fig. 1, suppose there are two identical hydrogen atoms A and B confined in two identical boxes A and B, respectively. Box A is carried by Alice who sets coordinate $(t,\vec{x})$ and Box B is carried by Bob who sets coordinate $(t',\vec{x}')$. In Alice's $(t,\vec{x})$ coordinate the Hamiltonian of the hydrogen atom A is $H_A(\vec{x})$, the ground and the first excited states are $\phi_1^A(t,\vec{x})$ and $\phi_2^A(t,\vec{x})$, respectively. Similarly, the atom B measured by Bob has $H_B(\vec{x}')$, $\phi_1^B(t',\vec{x}')$ and $\phi_2^B(t',\vec{x}')$. Here each of the two boxes can refer to an atomic clock device. We can denote $g_A^{\mu\nu}(t,\vec{x})$ as the space-time metric measured in Alice's $(t,\vec{x})$ coordinate of the location where Box A occupies. Now let us consider the case that $g_A^{\mu\nu}(t,\vec{x})$ is independent of time $t$, and the box has a very small size (it can be comparable with the size of atoms) such that the metric can be approximated as a constant within the size of the box, that is, we can define $g_A^{\mu\nu} \equiv g_A^{\mu\nu}(t,\vec{x})$ where $g_A^{\mu\nu}$ is a constant within the size of Box A measured in $(t,\vec{x})$ coordinate (the local space-time curvature within the box is ignored under this approximation). Now if Alice wants to describe the states of atoms and the interactions between atoms with the background fields inside of Box A, she needs to start from the local Lagrangian density of QED in curved space-time written as [24]

$$L_A = \bar{\psi}_A(t,\vec{x})[\gamma_A^\mu \nabla_\mu + m_0]\psi_A(t,\vec{x}) + e_0 j_A^\mu(t,\vec{x}) A_\mu^A(t,\vec{x}) - \frac{1}{4} F_{\mu\nu}^A(t,\vec{x}) F_A^{\mu\nu}(t,\vec{x}), \qquad (2.17)$$

where notation $A$ in $\gamma_A^\nu$, $\psi_A$, $A_\mu^A$, $F_{\mu\nu}^A$ and $j_A^\mu$ just denote that these quantities are defined related with Box A. We also have $\gamma_A^\mu \gamma_A^\nu + \gamma_A^\nu \gamma_A^\mu = 2g_A^{\mu\nu}$, $j_A^\mu(t,\vec{x}) = \frac{1}{2}[\bar{\psi}_A, \gamma_A^\mu \psi_A]$, $\nabla^\nu F_{\mu\nu}^A(t,\vec{x}) \equiv \nabla^\nu[\nabla_\nu A_\mu^A(t,\vec{x}) - \nabla_\mu A_\nu^A(t,\vec{x})]$ and $\nabla_\mu$ (it can be replaced by $\partial_\mu$ since the matric is approximated as a constant within the size of Box A) denotes the covariant derivative. All these quantities depend on the local metric value $g_A^{\mu\nu}$. From Eq. (2.17), the local fields $\psi_A(t,\vec{x})$ and $A_\mu^A(t,\vec{x})$ confined in Box A obey the equations obtained from Euler-Lagrange equations as

$$\nabla_\mu F_A^{\mu\nu}(t,\vec{x}) = -e_0 j_A^\nu(t,\vec{x}), \tag{2.18a}$$

$$(\gamma_A^\mu \nabla_\mu + m_0)\psi_A(t,\vec{x}) = -e_0 \gamma_A^\mu \psi_A(t,\vec{x}) A_\mu^A(t,\vec{x}). \tag{2.18b}$$

Indeed, the states of atoms and the interactions between atoms with background fields in Box A can be theoretically derived from Eq. (2.17) (here we only consider QED phenomena) with the local metric $g_A^{\mu\nu}$. Therefore, Alice can obtain the atomic transition frequency between states $\phi_1^A(t,\vec{x})$ and $\phi_2^A(t,\vec{x})$ as $\omega_A$. Now let Bob who is carrying Box B starts to move with a velocity $v_B(t)$ measured by Alice as shown in Fig. 1, we can also ask ourselves how to derive the states of atom B measured by Bob. Therefore, if Bob wants to describe the QED phenomena inside of Box B, he can set a coordinate system $(t',\vec{x}')$ and the metric where Box B occupies can be written as $g_B'^{\mu\nu}(\vec{x}')$. Similarly, he can apply the approximation as $g_B'^{\mu\nu} \equiv g_B'^{\mu\nu}(\vec{x}')$ where $g_B'^{\mu\nu}$ is a constant within the size of Box B measured in the coordinate $(t',\vec{x}')$. It is worth noting that $g_B'^{\mu\nu}$ is not the metric value measured by Alice using coordinate $(t,\vec{x})$. Then the Lagrangian density that Bob uses to describe the QED phenomena inside of Box B can be written as

$$L_B' = \bar{\psi}_B'(t',\vec{x}')[\gamma_B'^\mu \nabla_\mu' + m_0]\psi_B'(t',\vec{x}') + e_0 j_B'^\mu(t',\vec{x}') A_B'^\mu(t',\vec{x}') - \frac{1}{4} F_B'^{\mu\nu}(t',\vec{x}') F_{\mu\nu}'^B(t',\vec{x}'), \tag{2.19}$$

where $\gamma_B'^\mu \gamma_B'^\nu + \gamma_B'^\nu \gamma_B'^\mu = 2 g_B'^{\mu\nu}$, $\nabla'^\nu F_{\mu\nu}'^B(t',\vec{x}') \equiv \nabla'^\nu[\nabla_\nu' A_\mu'^B(t',\vec{x}') - \nabla_\mu' A_\nu'^B(t',\vec{x}')]$, $j_B'^\mu(t',\vec{x}') = \frac{1}{2}[\bar{\psi}_B', \gamma_B'^\mu \psi_B']$, and $\nabla_\mu'$ denotes the covariant derivative in $(t',x')$ coordinate. Compare Eq. (2.19) with Eq. (2.17), we notice that these two Lagrangians have the same mathematical structure, in other words, there is no additional effect that is caused by the motion or acceleration of Box B added into Eq. (2.19). This is because we assumed that the acceleration has no influence on the QFT physical phenomena as stated by MH. For a constant velocity $v_B$, we all agree that the Lagrangian in Box B is Eq. (2.19) by the one of the postulates in special relativity, here we just generalize this postulate to any non-inertial frames (this generalization only applies to QFT physics), later on we will see such generalization will produce the clock hypothesis. To be specific, recall that Newtonian physics takes a different mathematical form in non-inertial reference frames with an acceleration $a_B$, i.e., $F = m(a - a_B)$ in non-inertial reference frames which takes a different mathematical form compared with $F = ma$ in inertial frame. On the contrary with Newtonian physics, MH states that acceleration does not change the mathematical form of quantum field theory. We notice that the QED physics inside of Box B is completely determined by the local metric $g_B'^{\mu\nu}$ measured by Bob, and all the differences between the QED physics in two boxes are caused by the difference between the values of $g_A^{\mu\nu}$ and $g_B'^{\mu\nu}$. The field solutions inside of Box B can also be given as

$$\nabla'_\mu F_B'^{\mu\nu}(t',\vec{x}') = -e_0 j_B'^{\nu}(t',\vec{x}'), \tag{2.20a}$$

$$(\gamma_B'^{\mu}\nabla'_\mu + m_0)\psi'_B(t',\vec{x}') = -e_0\gamma_B'^{\mu}\psi'_B(t',\vec{x}')A_\mu'^{B}(t',\vec{x}'). \tag{2.20b}$$

In general, if $g_B'^{\mu\nu} \neq g_A^{\mu\nu}$, the physics formulas in Box B (such as $\psi'_B$, $j_B'^{\mu}$ and $A_\mu'^{B}$) will have different mathematical forms compared with that in Box A. As a result, the energy levels of the hydrogen atom may become different in the two boxes, that is $\omega'_B \neq \omega_A$ where $\omega'_B$ measured in Bob's $(t',\vec{x}')$ coordinate is the transition frequency between states $\phi_1'^{B}(t',\vec{x}')$ and $\phi_2'^{B}(t',\vec{x}')$ of atom B, this indicates that the periods of clock A and clock B which are built by the atoms inside of two the boxes will be different, i.e., the coordinate transformation between $(t,\vec{x})$ and $(t',\vec{x}')$ does not satisfy CTMPC. In order to know the physics formulas in Box B without repeating all the calculations done by Alice who has obtained physics formulas in Box A, Bob can set his coordinate $(t',\vec{x}')$ to satisfy

$$g_B'^{\mu\nu} = g_A^{\mu\nu}. \tag{2.21}$$

In other words, Bob can set his coordinate $(t',\vec{x}')$ such that the metric value of Box B measured in $(t',\vec{x}')$ coordinate equals to the metric value of Box A measured in $(t,\vec{x})$ coordinate, this can be done by adjusting the value of $(t',\vec{x}')$ (please see the sections IV and V for examples). Therefore, we get $\gamma_B'^{\mu} = \gamma_A^{\mu}$, then Eq. (2.19) will become the same as Eq. (2.17), and the field solutions of Eq. (2.20) will have the same mathematical forms compared with the field solutions of Eq. (2.18). That is, $\psi'_B(T,\vec{X}) = \psi_A(T,\vec{X})$ and $A_B'^{\mu}(T,\vec{X}) = A_A^{\mu}(T,\vec{X})$ where $(T,\vec{X})$ are any given time-space values. As a result, all the physical formulas and equations in the two boxes deducted from QED physics will become the same. In Bob's $(t',\vec{x}')$ coordinate, the Hamiltonian of the atom B is $H'_B = H_A$, the time-independent ground and the first excited states are $\phi_1'^{B} = \phi_1^{A}$ and $\phi_2'^{B} = \phi_2^{A}$, respectively. The energy levels of the two atoms are also the same, i.e., $\omega'_B = \omega_A$, therefore, CTMPC is satisfied.

At this stage, Bob knows that he can find a coordinate system $(t',\vec{x}')$ that both satisfies CTMPC and Eq. (2.21), and the time $t'$ indeed is the time displayed by clock B. As we can conclude: for QFT in curved space-times, any coordinate transformation that satisfies Eq. (2.21), which is named CTMLM, is one CTMPC.

Next let us look at an example to illustrate in details how CTMLM in Minkowski space-time is one CTMPC with atomic clocks. For Fig. 1 in Minkowski space-time, the metric is given as

$$g_A^{\mu\nu}(t,\vec{x}) = \begin{pmatrix} -1 & 0 \\ 0 & \vec{1} \end{pmatrix}. \tag{2.22}$$

In Box A, the QED Lagrangian density of Eq. (2.17) becomes

$$L = \bar{\psi}(t,\vec{x})[i\gamma^{\mu}(\partial_{\mu} + ie_0 A_{\mu}) - m_0]\psi(t,\vec{x}) - \frac{1}{4}F_{\mu\nu}(t,\vec{x})F^{\mu\nu}(t,\vec{x}). \tag{2.23}$$

The Hamiltonian of the hydrogen atom A can be deducted from QED physics as [1]

$$H_A = -\frac{1}{2m}\nabla^2 - \frac{e^2}{|\vec{x}|} \tag{2.24}$$

and the energy levels are

$$\omega_n^A = -\frac{me^4}{2n^2}. \tag{2.25}$$

The ground state of hydrogen atom A is $\phi_1^A(\vec{x})\exp(-i\omega_1^A t)$. If Bob who is moving with an arbitrary velocity $\vec{v}_B(t)$ [ $|\vec{v}_B(t)| < 1$ ] sets a coordinate system $(t',\vec{x}')$ that maintains the Minkowski metric, i.e. $g_B'^{\mu\nu}(t',\vec{x}') = g_A^{\mu\nu}(t,\vec{x})$, then he will use the below Lagrangian density to describe the QED phenomena inside of Box B as

$$L' = \bar{\psi}(t',\vec{x}')[i\gamma^{\mu}(\partial'_{\mu} + ie_0 A_{\mu}) - m_0]\psi(t',\vec{x}') - \frac{1}{4}F_{\mu\nu}(t',\vec{x}')F^{\mu\nu}(t',\vec{x}'). \tag{2.26}$$

Therefore, the two Lagrangians have the same mathematical structure, the only difference between them is the value of space-time parameters, and the relationship between the values of $(t',\vec{x}')$ and $(t,\vec{x})$ can be deducted from Eq. (2.21) (we will do this in following sections). The Hamiltonian of the hydrogen atom B written in $(t',\vec{x}')$ coordinate can be deducted from Eq. (2.26) as

$$H_B = -\frac{1}{2m}\nabla'^2 - \frac{e^2}{|\vec{x}'|}. \tag{2.27}$$

Comparing Eq. (2.27) with Eq. (2.24), we can see that the atom B has a ground state expressed by $\phi_1^A(\vec{x}')\exp(-i\omega_1^A t')$. In other words, the expression of the quantum states of the atom B can be obtained from the corresponding quantum states of the atom A by replacing $(t,\vec{x})$ with $(t',\vec{x}')$. And the energy levels $\omega_n'^B$ of atom B are exactly the same as $\omega_n^A$ expressed by Eq. (2.25). Therefore, the frequency $\omega'$ measured in coordinate $(t',\vec{x}')$ of the photon emitted by the atomic transition between the first excited state and the ground state of atom B equals to the frequency $\omega$ measured in coordinate $(t,\vec{x})$ of the emitted photon caused by the same transition of the atom A. Thus, the two clocks have the same period value and CTMPC is satisfied, the time $t'$ in coordinate $(t',\vec{x}')$ is indeed the time displayed by clock B. Note that in order to make $\omega'=\omega$, this $\omega'$ must be measured in coordinate $(t',\vec{x}')$ which maintains the local metric value expressed by Eq. (2.21). In the case of $g_B'^{\mu\nu} \neq g_A^{\mu\nu}$, the mathematical form of states of atom B will not be the same as states of atom A, thus the energy levels measured in the corresponding coordinate will not be the same either. In this paper, the purpose of all the above discussions from Eq. (2.17) to Eq. (2.27) is to lay a solid theoretical background for the reason why CTMLM in any curved space-times is one type of CTMPC. In the following sections, we will show how we can use CTMLM to study the time dilation and Doppler effect in curved space-time. Before that, in order to make our discussions more clearly, we can look at another example.

There are two identical atomic clocks A (carried by Alice) and B (carried by Bob) in Minkowski space-time, Bob is moving with a constant speed $v_B$ measured by Alice using $(t,x)$ coordinate. Suppose the atomic transition in clock A generates a plane wave $e^{i(\omega t-kx)}$ measured by Alice in coordinate $(t,x)$ and the atomic transition in clock B generates another plane wave $e^{i(\omega' t'-k' x')}$ measured by Bob in coordinate $(t',x')$. We know that for a constant speed $v_B$ the Lorentz transformation is given as

$$t' = \gamma(t - v_B x), \tag{2.28a}$$

$$x' = \gamma(x - v_B t), \tag{2.28b}$$

which maintains the local metric value with $\gamma = 1/\sqrt{1-v_B^2}$, i.e., Eq. (2.28) satisfy Eq. (2.21). As discussed before, the quantum states measured in coordinate $(t',\vec{x}')$ of the atoms in clock B are the same as the quantum states measured in coordinate $(t,\vec{x})$ of the atoms in clock A. Therefore, the frequency $\omega'$ in $e^{i(\omega' t'-k' x')}$ is equal to $\omega$ in $e^{i(\omega t-kx)}$, i.e., $\omega'=\omega$. Therefore, we can obtain the cycles completed by $e^{i(kx-\omega t)}$ as $n=\dfrac{t}{T}=\dfrac{\omega t}{2\pi}$ and the cycles completed by $e^{i(k' x'-\omega' t')}$ as $n'=\dfrac{t'}{T'}=\dfrac{\omega' t'}{2\pi}$, we can see that $\dfrac{t'}{t}=\dfrac{n'}{n}$ due to $\omega'=\omega$. Thus, Eq. (2.28) indeed shows the correct relation between the times displayed by the two clocks.

In conclusion, CTMPC is introduced in this section and its significance is discussed. We proved that CTMPC agrees with the time measured by clocks, as a result, it is a correct coordinate transformation in curved space-times. For QFT in curved space-times, we demonstrate that CTMLM as indicated by Eq. (2.21) is one CTMPC. Moreover, we show that the mathematical forms of physical formulas can be maintained for CTMLM. Therefore, we do not have to repeat all the calculations to obtain the physical formulas in another coordinate system, the only difference of the physical formulas between the two coordinates is the value of space-time parameters ($t \to t', x \to x'$), and the time difference between $t'$ and $t$ will be shown by the two clocks. This is an advantage of setting a new coordinate $(t',x')$ that maintains the local metric value, soon it also will be shown as a powerful tool to study the Doppler effect in following sections III, IV, V.

## III. Time Dilation and Doppler Effect with Time-Dependent Velocity in Minkowski Space-Time

The two dimensional Minkowski space-time has the metric

$$g^{\mu\nu}(\tilde{x})=\begin{pmatrix}-1 & 0\\ 0 & 1\end{pmatrix} \tag{3.1}$$

with the defined symbol $\tilde{x}^0\equiv t$ and $\tilde{x}^1\equiv x$. Suppose that we have two identical atomic clocks A and B rest at $x=0$ which are carried by Alice and Bob, respectively. At time $t=0$, Bob starts to move away from Alice and he sets a new coordinate system $(t',x')$ to measure the physical events observed by himself. Now we can write the coordinate transformation that maintains the metric value as

$$dt'=\cosh f(\tilde{x})dt-\sinh f(\tilde{x})dx, \tag{3.2a}$$

$$dx'=-\sinh f(\tilde{x})dt+\cosh f(\tilde{x})dx \tag{3.2b}$$

in which $f(\tilde{x})$ can be any arbitrary functions of $t$ and $x$. Similar with the Lorentz transformation, Eq. (3.2) can be used to transform $(dt,dx)$ between any two nearby physical events measured by Alice into $(dt',dx')$ between the same two physical events measured by Bob. The velocity of Bob $v_B(t)$ measured by Alice can be obtained by demanding $dx'=0$ in Eq. (3.2b), we thus get the velocity of Bob as

$$v_B(t)\equiv\frac{dx}{dt}=\tanh f(\tilde{x})\,. \tag{3.3}$$

The first order partial derivatives can be obtained from Eq. (3.2) as

$$\frac{\partial t'}{\partial t}=\frac{\partial x'}{\partial x}=\cosh f(\tilde{x})\,, \tag{3.4a}$$

$$\frac{\partial t'}{\partial x}=\frac{\partial x'}{\partial t}=-\sinh f(\tilde{x})\,. \tag{3.4b}$$

The metric $g'^{\rho\sigma}(\tilde{x}')$ in the $(t',x')$ coordinate can be given by Eq. (2.13). We can verify that $g'^{\rho\sigma}(\tilde{x}')=g^{\rho\sigma}(\tilde{x})$ shown as Eq. (3.1), that is, the metric is unchanged after the transformation and it satisfies CTMLM. Based on the argument of CTMPC and CTMLM, since the value of the metric are the same at the spatial location where the two clocks occupy, the frequency of the emissions generated by the atomic transitions inside of the clocks will be the same and the two clocks will thus have the same period, therefore, the relationship between $dt$ and $dt'$ shown by Eq. (3.2a) is indeed the correct relationship for the displayed time of the two clocks.

Note that different transformation function $f(\tilde{x})$ in Eq. (3.2) corresponds to different $v_B(t)$, $f(\tilde{x})$ for the Lorentz transformation with constant $v_B$ can be worked out. The transformed coordinate $\tilde{x}'(\tilde{x})$ as functions of $\tilde{x}$ can be obtained by integration of Eq. (3.2) along a path, that is,

$$t'(\tilde{x}) \equiv \int_{path} dt' = \int_{path} \cosh f(\tilde{x})dt - \sinh f(\tilde{x})dx, \tag{3.5a}$$

$$x'(\tilde{x}) \equiv \int_{path} dx' = \int_{path} -\sinh f(\tilde{x})dt + \cosh f(\tilde{x})dx. \tag{3.5b}$$

However, in order to get an integration-path-independent (IPI) function $\tilde{x}'(\tilde{x})$ (the integration value only depends on two end points of a path), the second order partial derivatives need to satisfy the following relationships

$$\frac{\partial^2 t'}{\partial x \partial t} = \frac{\partial^2 t'}{\partial t \partial x}, \tag{3.6a}$$

$$\frac{\partial^2 x'}{\partial x \partial t} = \frac{\partial^2 x'}{\partial t \partial x}. \tag{3.6b}$$

By substituting Eq. (3.6) into Eq. (3.4), we get

$$\frac{\partial \cosh f(\tilde{x})}{\partial x} = -\frac{\partial \sinh f(\tilde{x})}{\partial t}, \tag{3.7a}$$

$$\frac{\partial \cosh f(\tilde{x})}{\partial t} = -\frac{\partial \sinh f(\tilde{x})}{\partial x}. \tag{3.7b}$$

One solution can be given as $f(\tilde{x}) = \Omega$ in which $\Omega$ is a constant. We can recognize that this solution corresponds to the Lorentz transformation with a constant velocity, the velocity of Bob is $v_B = \tanh \Omega$. Therefore, the Lorentz transformation is well-known as an isometry of Minkowski metric.

We may also wonder what happens if the velocity of Bob $v_B(t) = \tanh f(\tilde{x})$ is not a constant. Since for a general function $f(\tilde{x})$ that does not satisfy Eq. (3.7), we cannot get the coordinates $\tilde{x}'(\tilde{x})$ as IPI functions of $\tilde{x}$. Nevertheless, we can still integrate along a certain path to obtain the integration-path-dependent (IPD) functions $\tilde{x}'(\tilde{x})$. Next we will look at one example to illustrate this point and later on we will see how it is related to the twin paradox.

As shown in Fig. 2, the coordinate system $(t, x)$ is set by Alice who carries the clock A. At $t = 0$, Bob who carries the clock B starts to move with a constant positive acceleration $a \equiv \dfrac{d^2x}{dt^2} = a_B$ ; at time $t = t_1$ (note that this $t = t_1$ is measured by clock A and $a_B t_1 < 1$), Bob starts to decelerate with $a = -a_B$ ; at $t = 2t_1$ , Bob would start to move back with $a = -a_B$ ; finally, at $t = 3t_1$ Bob starts to decelerate with $a = a_B$ and he meets Alice at $t = 4t_1$ . As can be seen from Fig. 2, during the time period $0 \le t \le 4t_1$ , Alice follows the path I and Bob follows the path II, they meet at $(t = 4t_1, x = 0)$ .

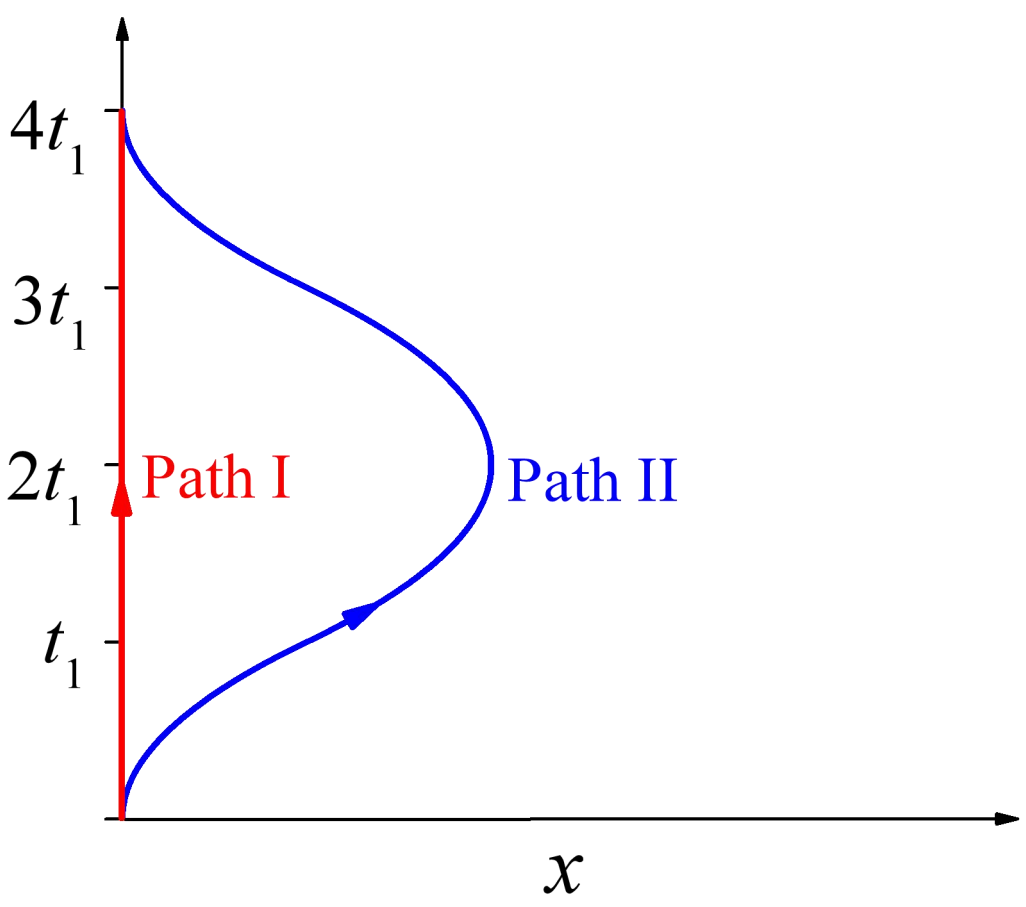

FIG. 2. Sketch of the two Paths measured in Alice's coordinate $(t, x)$ . The red line indicates Path I that Alice follows and the blue curve is Path II that Bob travels, Bob turns back at $t = 2t_1$ and they meet at $(t = 4t_1, x = 0)$ .

In order to calculate the time difference between the two clocks when they meet by using Eq. (3.5a), we can treat the ticking of clock B as physical events and all these events occur along path II. Therefore, we can integrate along path II to obtain the total time of the events measured by Alice in coordinate $(t, x)$ , and the total time of the events measured by Bob is defined as $t'_{II}(\tilde{x})$ in coordinate $(t', x')$, this $t'_{II}(\tilde{x})$ is also what the clock B shows when they meet. Note that because Eq. (3.2) is the coordinate transformation of the same events observed by different observers, we cannot just perform the integrations along path I for clock A and along path II for clock B, and then compare the time difference between the two different paths.

For Alice in coordinate $(t, x)$, if we integrate along the path II, we get

$$t = \int_{path\ II} dt \ , \tag{3.8a}$$

$$x = \int_{path\ II} dx. \tag{3.8b}$$

Note that $\int_{path\,II} dt$ is the total time of the events (ticking of clock B) occurring along path II measured by Alice. Since path I combined with path II forms a loop, we also know that in coordinate $(t, x)$, the integration is path-independent, that is,

$$\int_{path\,II} dt = \int_{path\,I} dt = 4t_1\,, \tag{3.9a}$$

$$\int_{path\,II} dx = \int_{path\,I} dx = 0. \tag{3.9b}$$

Note that $\int_{path\,I} dt$ in Eq. (3.9a) is the time that clock A shows when they meet. Therefore, clock A whose motion follows the path I can also be used to measure the total time of the events (ticking of clock B) occurring along path II. For coordinate $(t', x')$, if we integrate along path II, according to Eq. (3.5), we obtain (see Appendix A)

$$t'_{II}(\tilde{x}) = \int_{path\,II} \cosh f(\tilde{x})dt - \sinh f(\tilde{x})dx = \frac{2}{a_B}\arcsin a_B t_1 + 2t_1\sqrt{1 - a_B^2 t_1^2}\,, \tag{3.10a}$$

$$x'_{II}(\tilde{x}) = \int_{path\,II} -\sinh f(\tilde{x})dt + \cosh f(\tilde{x})dx = 0. \tag{3.10b}$$

Since $t'_{II}(\tilde{x})$ is the time shown by clock B when they meet and we can see that $t'_{II}(\tilde{x}) < 4t_1$, therefore, the travelling clock runs slower. At this stage, we notice that the results of Eq. (3.8) and Eq. (3.10) are both integrations of path II observed by Alice and Bob, respectively. By symmetry, we can ask ourselves: what if we treat the ticking of clock A as the physical events that we want to measure? Then we need to integrate along path I, Eq. (3.8) remain the same since the result is path-independent as shown by Eq. (3.9). For coordinate $(t', x')$ set by Bob, he obtains (see Appendix A)

$$t'_{I}(\tilde{x}) = \int_{path\,I} \cosh f(\tilde{x})dt - \sinh f(\tilde{x})dx = \frac{4}{a_B}\arcsin a_B t_1, \tag{3.11a}$$

$$x'_{I}(\tilde{x}) = \int_{path\,I} -\sinh f(\tilde{x})dt + \cosh f(\tilde{x})dx = 0. \tag{3.11b}$$

We can see $t'_{I}(\tilde{x}) \neq t'_{II}(\tilde{x})$, this is as expected actually since we already discussed earlier that $(t', x')$ will be IPD functions if $v_B(t)$ is not a constant. Therefore, clock B whose motion follows the path II cannot be used to measure the total time of the physical events (ticking of clock A) occurring along path I since the time shown by clock B is $t'_{II}(\tilde{x})$ when they meet at $t = 4t_1$. This fact can be used to explain the twin-paradox: we can only integrate along path II (followed by the moving twin) to calculate the time difference between the twins.

Next we can take a look at another example. Two identical atomic clocks are carried by Alice and Bob, respectively. They are initially at rest and Alice sets a coordinate system $(t, x)$ while Bob sets another coordinate $(t',x')$ which maintains the local metric given by Eq. (3.2). At $t=0$, Bob starts to move with constant acceleration $\alpha$ measured by Alice's $(t,x)$ coordinate satisfying

$$f(\tilde{x})=\alpha\tau\ . \tag{3.12}$$

In Eq. (3.12), $\tau$ is the proper time measured by Alice and

$$t(\tau)=\frac{1}{\alpha}\sinh(\alpha\tau)\ , \tag{3.13a}$$

$$x(\tau)=\frac{1}{\alpha}\cosh(\alpha\tau)\ . \tag{3.13b}$$

The velocity of Bob measured by Alice can be expressed as

$$v_B(t)\equiv\frac{dx}{dt}=\tanh(\alpha\tau)=\tanh[\text{arcsinh}(\alpha t)]=\frac{\alpha t}{\sqrt{1+\alpha^2t^2}}\ . \tag{3.14}$$

From Eq. (3.2a), the time difference between the two clocks can be given as

$$dt'=[\cosh f(\tilde{x})-v_B(t)\sinh f(\tilde{x})]dt=\frac{1}{\cosh\alpha\tau}dt=\sqrt{1-v_B^2(t)}dt \tag{3.15}$$

which is the time dilation with time-dependent velocity in Minkowski space-time, this result is consistent with the clock hypothesis. The scalar field $\psi(t,x)$ that satisfies Eq. (2.11) with Minkowski metric can be expressed as the superposition of plane waves

$$\psi(t,x)=\sum_k[a_k e^{-i(\omega t-kx)}+a_k^+e^{i(\omega t-kx)}] \tag{3.16}$$

with $\omega=k$ . In Bob's coordinate $(t',x')$ that maintains the local metric value, his Lagrangian describing QED physics is given by Eq. (2.25), therefore the modes can be obtained without any more calculations. As we argued in section II, this is a big advantage to maintain the local metric value, that is, the mathematical form for the physical formulas in coordinate $(t',x')$ is exactly the same as that in coordinate $(t,x)$, what left for Bob to do is just replacing $(t,x)$ in Eq. (3.16) with $(t',x')$ and obtain the modes in coordinate $(t',x')$ as

$$\psi(t',x')=\sum_{k'}[a_{k'}e^{-i(\omega't'-k'x')}+a_{k'}^+e^{i(\omega't'-k'x')}] \tag{3.17}$$

with $\omega'=k'$ . Now suppose that an atom carried by Alice emits a plane wave $e^{-i(\omega t-kx)}$ to Bob, Bob can decompose this plane wave in his coordinate $(t',x')$ as

$$e^{-i(\omega t-kx)}=\sum_{\omega'}[\alpha_{\omega\omega'}e^{-i(\omega't'-k'x')}+\beta_{\omega\omega'}e^{i(\omega't'-k'x')}] \tag{3.18}$$

in which $\alpha_{\omega\omega'}$ and $\beta_{\omega\omega'}$ are parameters to be fixed. For a trial solution, we require $\beta_{\omega\omega'}=0$, and $\alpha_{\omega\omega'}\neq 0$ only if $\omega'$ equals one certain value to be fixed, then we get

$$e^{-i(\omega t-kx)}=e^{-i(\omega' t'-k'x')}\Rightarrow \omega t-kx=\omega' t'-k'x'+2n\pi\,. \tag{3.19}$$

Therefore, after an infinitesimal time interval $dt$ we have

$$\omega dt-kdx=\omega' dt'-k'dx'\,. \tag{3.20}$$

In Bob's co-moving frame $(t',x')$, $dx'=0$ and we obtain

$$\omega'=\frac{\omega dt-kdx}{dt'}=\omega[1-v_B(t)]\frac{dt}{dt'}=\frac{\omega[1-v_B(t)]}{\sqrt{1-v_B^2(t)}}=\omega\sqrt{\frac{1-v_B(t)}{1+v_B(t)}} \tag{3.21}$$

in which $\frac{dt}{dt'}$ is given by Eq. (3.15) and $v_B(t)$ is given by Eq. (3.14). Therefore, the Doppler-shift with a time-dependent velocity is determined by the velocity at that instant. This conclusion is obtained in accordance with the clock hypothesis in special relativity. The relationship between the creation and annihilation operators in two coordinate systems can be given as

$$a_{k'}=a_k\,, \tag{3.22a}$$

$$a_{k'}^{+}=a_k^{+} \tag{3.22b}$$

with $k'=k\sqrt{\frac{1-v_B(t)}{1+v_B(t)}}$. Therefore, for Eq. (3.16) and Eq. (3.17) we can also verify

$$\psi(t,x)=\psi(t',x') \tag{3.23}$$

which agrees with Eq. (2.16).

## IV. Time Dilation and Doppler Effect with Time-Dependent Velocity in Schwarzschild Space-Time

For simplicity, we will only consider the coordinate transformations in $(t,r)$ directions for the Schwarzschild metric, therefore, the two dimensional Schwarzschild metric can be described by

$$g_{\mu\nu}(t,r)=\begin{pmatrix}-(1-\frac{2M}{r}) & 0\\ 0 & (1-\frac{2M}{r})^{-1}\end{pmatrix}. \tag{4.1}$$

Similarly as Minkowski space-time, if we want to find another coordinate system $(t',r')$ that maintains the metric form, the metric $g'_{\mu\nu}(t',r')$ is given as

$$g'_{\mu\nu}(t',r') = g_{\mu\nu}(t',r') = \begin{pmatrix} -(1-\frac{2M}{r'}) & 0 \\ 0 & (1-\frac{2M}{r'})^{-1} \end{pmatrix}. \tag{4.2}$$

The metric transformation between different coordinate systems can be expressed by

$$g_{\mu\nu} = \frac{\partial \tilde{x}'^{\rho}}{\partial \tilde{x}^{\mu}} \frac{\partial \tilde{x}'^{\sigma}}{\partial \tilde{x}^{\nu}} g'_{\rho\sigma}. \tag{4.3}$$

Therefore, we obtain

$$g_{00} = -(1-\frac{2M}{r}) = -(1-\frac{2M}{r'})(\frac{\partial t'}{\partial t})^2 + (1-\frac{2M}{r'})^{-1}(\frac{\partial r'}{\partial t})^2, \tag{4.4a}$$

$$g_{01} = g_{10} = 0 = -(1-\frac{2M}{r'})\frac{\partial t'}{\partial t}\frac{\partial t'}{\partial r} + (1-\frac{2M}{r'})^{-1}\frac{\partial r'}{\partial t}\frac{\partial r'}{\partial r}, \tag{4.4b}$$

$$g_{11} = (1-\frac{2M}{r})^{-1} = -(1-\frac{2M}{r'})(\frac{\partial t'}{\partial r})^2 + (1-\frac{2M}{r'})^{-1}(\frac{\partial r'}{\partial r})^2. \tag{4.4c}$$

Similarly as Minkowski space-time, if we want to find an IPI solution for $(t',r')$ as functions of $(t,r)$, the second order partial derivatives need to satisfy the following equations

$$\frac{\partial^2 t'}{\partial r \partial t} = \frac{\partial^2 t'}{\partial t \partial r}, \tag{4.5a}$$

$$\frac{\partial^2 r'}{\partial r \partial t} = \frac{\partial^2 r'}{\partial t \partial r}. \tag{4.5b}$$

In this section, we will not go further to find such IPI coordinate $(t',r')$, but we assume that the coordinate $(t',r')$ under study in this section is IPD. Meanwhile, we follow the convention that the coordinate system $(t,r)$ is set at spatial infinity $r \to \infty$. That is, if Alice who carries clock A is rest at $r \to \infty$, she can keep record of every physical event that occurs at location $(t,r)$, where $t$ is the time displayed by clock A and $r$ is the spatial distance from the gravitational source (note that $r$ is not the distance from Alice).

Now we consider two observers Alice who carries the atomic clock A and Bob who carries an identical clock B. Alice who sets the coordinate system $(t,r)$ is rest at $r_A \to \infty$ and $r_A$ is the location of clock A, Bob is located at $r = r_B$ and this $r_B$ is also measured in Alice's coordinate system. Bob himself sets another coordinate system $(t',r')$. Note that in general, Bob cannot find another IPI coordinate system given by Eq. (4.2) even if he is rest at $r = r_B$, this is because Eq. (4.5) do not have a general solution for the spatial translation from $r \to \infty$ to $r = r_B$. At time $t = 0$ and $t' = 0$, Bob starts to move with velocity $v_B(t) \equiv \frac{dr_B}{dt}$. In order to compare the two clocks, based on the argument of CTMPC and CTMLM, we must use the same metric value to describe clock B and clock A, that is, the components of the metric where clock B occupies measured by Bob equals to the components of metric where clock A occupies measured by Alice. Therefore, we have

$$g^{A}_{\mu\nu}(t, r_A) = g'^{B}_{\mu\nu}(t', r'_B), \tag{4.6}$$

where $g^{A}_{\mu\nu}(t, r_A)$ is given by Eq. (4.1) with $r$ replaced by $r_A$ and $g'^{B}_{\mu\nu}(t', r'_B)$ is the metric written in Bob's coordinate $(t', r')$, $r'_B$ is the spatial value of Bob's location measured in his coordinate $(t', r')$. Due to $r_A \to \infty$, we get

$$g'^{B}_{\mu\nu}(t', r'_B) = g^{A}_{\mu\nu}(t, r_A) = \begin{pmatrix} -1 & 0 \\ 0 & 1 \end{pmatrix}. \tag{4.7}$$

Therefore, the metric value at the position of clock A measured in Alice's coordinate $(t, r)$ equals to the metric value at the position of clock B measured in Bob's coordinate $(t', r')$. Thus, the frequency $\omega'$ of the wave emitted by the atom inside of clock B measured in Bob's coordinate $(t', r')$ equals to the transition frequency $\omega$ of clock A measured by Alice's coordinate $(t, r)$. As a result, the period of clock B measured in the coordinate $(t', r')$ will equal to that of clock A measured in the coordinate $(t, r)$, thus the time displayed by clock B can be given by $t'$ in Eq. (4.8). For the ticking event of clock B, according to Eq. (4.7), Eq. (4.4) can be rewritten as

$$-[1 - \frac{2M}{r_B(t)}] = -(\frac{\partial t'}{\partial t})^2 + (\frac{\partial r'}{\partial t})^2, \tag{4.8a}$$

$$0 = -\frac{\partial t'}{\partial t}\frac{\partial t'}{\partial r} + \frac{\partial r'}{\partial t}\frac{\partial r'}{\partial r}, \tag{4.8b}$$

$$[1 - \frac{2M}{r_B(t)}]^{-1} = -(\frac{\partial t'}{\partial r})^2 + (\frac{\partial r'}{\partial r})^2. \tag{4.8c}$$

In practice, Bob does not have to specify the metric values at other locations with $r' \neq r'_B$, since Eq. (4.7) is all the information we need to know about the local experiment performed at location $r'_B$. However, if Bob wants to set the full coordinate system $(t', r')$ that extends to the entire space-time with the metric Eq. (4.2), he needs to demand $r'_B = r_A$ based on Eq. (4.6) and Eq. (4.2), that is, $r_B' \to \infty$ . For convenience only, let us just assume that Bob himself indeed sets such coordinate system with the metric given by Eq. (4.2) (note that such coordinate system would be IPD). In Eq. (4.8), $r_B(t)$ is the spatial location of clock B measured by Alice and it depends on $t$ since Bob is moving with velocity $v_B(t) \equiv \frac{dr_B(t)}{dt}$. We can obtain the time relation between the two clocks as (see Appendix B)

$$dt' = \sqrt{\frac{[1-\frac{2M}{r_B(t)}]^2 - v_B^2(t)}{1-\frac{2M}{r_B(t)}}}dt\,. \tag{4.9}$$

Upon integration along the path that clock B travels, we get $t'$ and $t$ as the times displayed by clock B and clock A, respectively. As we can see, the time difference is determined by the location and the velocity of Bob measured by Alice at that instant, this result agrees with the clock hypothesis.

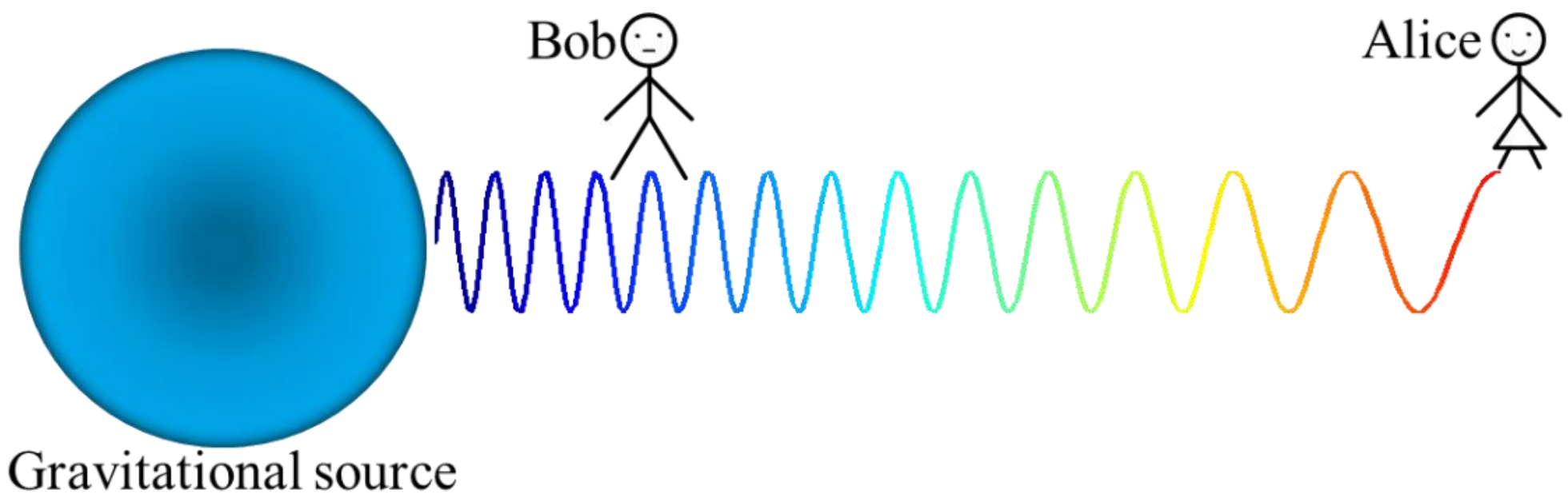

FIG. 3. Sketch of the Doppler effect near the gravitational source. Alice who is at spatial infinity sets a coordinate system $(t,r)$ and Bob who is located at $r_B(t)$ measured by Alice sets a coordinate system $(t',r')$ . Alice sends a wave with frequency $\omega$ measured in coordinate $(t,r)$ to Bob, the frequency of this wave becomes $\omega'$ measured in Bob's $(t',r')$ coordinate. The frequency change from $\omega$ to $\omega'$ is due to the switching of the reference frames from $(t,r)$ to $(t',r')$ .

The mode solution of Eq. (2.11) with Schwarzschild metric can be obtained as [25–27]

$$\psi(\tilde{x}) = \sum_{\omega,l,m} [a_{\omega,l,m} R_{\omega,l}(r) Y_{l,m}(\vartheta,\varphi) e^{-i\omega t} + a^+_{\omega,l,m} R^*_{\omega,l}(r) Y^*_{l,m}(\vartheta,\varphi) e^{i\omega t}] \tag{4.10}$$

in which $Y_{l,m}(\vartheta,\varphi)$ is the Bessel function with mode number $l,m$ , $R_{\omega,l}(r)$ is a function of $r$ that satisfies a second order differential equation and $a_{\omega,l,m}$ is the annihilation operator for mode $\omega,l,m$ . Since we are dealing with two-dimensional coordinate transformation in this paper, we will drop symbols $l,m$ and $Y_{l,m}(\vartheta,\varphi)$ for all subsequent discussions. Then Eq. (4.10) can be written as

$$\psi(t,r)=\sum_{\omega}[a_{\omega}R_{\omega}(r)e^{-i\omega t}+a_{\omega}^{+}R_{\omega}^{*}(r)e^{i\omega t}]\,. \tag{4.11}$$

According to Bob's coordinate $(t',r')$ with the metric given by Eq. (4.2), Eq. (4.11) can be written as

$$\psi(t',r')=\sum_{\omega'}[a'_{\omega'}R_{\omega'}(r')e^{-i\omega't'}+a'^{+}_{\omega'}R_{\omega'}^{*}(r')e^{i\omega't'}]\,. \tag{4.12}$$

As shown by Fig. 3, Alice sends a wave $R_{\omega}(r)e^{-i\omega t}$ with frequency $\omega$ to Bob, then Bob decomposes this wave $R_{\omega}(r)e^{-i\omega t}$ into the modes with positive and negative frequencies in coordinate $(t',r')$ as

$$R_{\omega}(r)e^{-i\omega t}=\sum_{\omega'}[\alpha_{\omega\omega'}R_{\omega'}(r')e^{-i\omega't'}+\beta_{\omega\omega'}R_{\omega'}^{*}(r')e^{i\omega't'}] \tag{4.13}$$

in which $\alpha_{\omega\omega'}$ and $\beta_{\omega\omega'}$ are parameters to be fixed. When the wave reaches Bob at location $r_B(t)$, this location in Bob's $(t',r')$ coordinate is $r'_B\to\infty$. Since $r'_B\to\infty$, the metric becomes Minkowski shown by Eq. (4.7), and thus the wave $R_{\omega'}(r')e^{-i\omega't'}$ in Eq. (4.13) becomes $e^{-i(\omega't'+k'r')}$ with $\omega'=k'$. Note that since the wave is traveling with negative velocity, we write the plane wave as $e^{-i(\omega't'+k'r')}$ instead of $e^{-i(\omega't'-k'r')}$. Therefore, we have

$$R_{\omega}(r_B)e^{-i\omega t}=\sum_{\omega'}[\alpha_{\omega\omega'}e^{-i\omega'(t'+r'_B)}+\beta_{\omega\omega'}e^{i\omega'(t'+r'_B)}]\,. \tag{4.14}$$

After an infinitesimal time interval $dt$, along Bob's co-moving frame $dr'_B=0$, we have

$$R_{\omega}(r_B+v_B dt)e^{-i\omega(t+dt)}=\sum_{\omega'}[\alpha_{\omega\omega'}e^{-i\omega'(t'+dt'+r'_B)}+\beta_{\omega\omega'}e^{i\omega'(t'+dt'+r'_B)}]\,. \tag{4.15}$$

Expanding Eq. (4.15) to first order, we obtain

$$e^{-i\omega t}v_B\frac{dR_{\omega}(r)}{dr}\Big|_{r=r_B}-i\omega e^{-i\omega t}R_{\omega}(r_B)=\sum_{\omega'}i\omega'[-\alpha_{\omega\omega'}e^{-i\omega'(t'+r'_B)}+\beta_{\omega\omega'}e^{i\omega'(t'+r'_B)}]\frac{dt'}{dt} \tag{4.16}$$

in which $\dfrac{dt'}{dt}$ is given by Eq. (4.9). Note that unlike Eq. (3.18) in Minkowski space-time, $\beta_{\omega\omega'}$ in Eq. (4.14) may not, in general, all vanish, it means that a mode with only positive frequency measured by Alice may become modes involving negative frequencies measured by Bob. For Minkowski space-time, $\beta_{\omega\omega'}$ always vanish since the plane wave $\dfrac{1}{e^{-ikr}}\dfrac{de^{-ikr}}{dr}$ is a pure imaginary number ; on the contrary, $\frac{1}{R_\omega(r_B)}\frac{dR_\omega(r_B)}{dr}$ in Eq. (4.16) is not always an imaginary number, thus we cannot ascertain to get a real solution $\omega'$ if all $\beta_{\omega\omega'}$ equal to zero, usually this phenomenon is interpreted as particle-creation in curved space-time [28–30]. That is, the modes of Eq. (4.11) in Alice's coordinate system are super-positioned states measured in Bob's coordinate, as is shown by Eq. (4.14). At this stage, one may wonder why Bob needs to set his coordinate that maintains the local metric at Alice's location when he measures the particle sent by Alice. The reason is specified as below.

Alice's atom is far away from the gravitational source when emitting the wave and this wave can be given as plane wave $e^{i(kx-\omega t)}$ at Alice's location. Upon arriving at Bob's location, this wave becomes $R_\omega(r_B)e^{-i\omega t}$ given by Eq. (4.11). Meanwhile, the physical phenomenon (atom emits a photon) can be described using Eq. (2.17) with the Minkowski metric. Bob who is carrying an identical atom needs to reproduce a reverse physical phenomenon (a photon absorbed by the atom), thus, he needs to put his atom on the equal footing with Alice's atom. That is, he knows that his atom will absorb a plane wave $e^{-i(\omega't'-k'x')}$ with the same frequency as the wave $e^{-i(\omega t-kx)}$ emitted by Alice's atom, i.e., $\omega=\omega'$. Therefore, he can set his local coordinate system with the same Minkowski metric. However, the wave sent by Alice becomes $R_\omega(r_B)e^{-i\omega t}$ after the propagation all the way from Alice to Bob, and this wave $R_\omega(r_B)e^{-i\omega t}$ is still measured in Alice's coordinate system with the metric given by Eq. (4.1) at Bob's location. Thus, Bob needs to decompose this mode in his coordinate system in order to find out whether this wave can be absorbed by his atom or not. Thus, the wave $R_\omega(r_B)e^{-i\omega t}$ need to be decomposed into a range of plane waves given by Eq. (4.14). In this paper, we will not go further to solve Eq. (4.14) and Eq. (4.16) to obtain $\alpha_{\omega\omega'}$ and $\beta_{\omega\omega'}$, instead, we will look at some simpler cases.

First let us look at a case with $v_B=0$, Eq. (4.16) becomes

$$-i\omega R_\omega(r_B)e^{-i\omega t}=\sum_{\omega'} i\omega'[-\alpha_{\omega\omega'}e^{-i\omega'(t'+r_B')}+\beta_{\omega\omega'}e^{i\omega'(t'+r_B')}]\frac{dt'}{dt}. \tag{4.17}$$

We apply a trial solution with $\beta_{\omega\omega'}=0$ (all the $\beta_{\omega\omega'}$ vanish) and $\alpha_{\omega\omega'}$ is non-zero only at one certain $\omega'$ to be fixed, thus combine Eq. (4.17) with Eq. (4.14), we have

$$\omega'=\frac{\omega}{\sqrt{1-\frac{2M}{r_B}}} \tag{4.18}$$

which is the gravitational blue-shift measured by Bob who is rest at $r_B$. In this case, the relationship between $a_{\omega'}$ in Eq. (4.12) and $a_\omega$ in Eq. (4.11) can be given as

$$a_{\omega'}=a_\omega \tag{4.19}$$

with $\omega'$ given by Eq. (4.18). Note that the frequency $\omega$ of the wave $R_\omega(r)e^{-i\omega t}$ does not change during the propagation measured in Alice's coordinate $(t,r)$, the frequency shift given by Eq. (4.18) is due to the switching of the reference frame from $(t,r)$ to $(t',r')$, this interpretation agrees with Einstein's equivalence principle: the changing of the photon frequency in the gravitational field is equivalent with the changing frequency in accelerating reference frames [31]. One may find that such interpretation contradicts with the energy-conservation argument of photons (the energy of the photon together with the gravitational potential is conserved, since the potential energy is altered at different locations, the photon energy will be shifted accordingly). However, this energy-conservation argument is purely classical and it originates from the observation of macroscopic objects falling in the gravitational field. We can clarify this issue here by comparing the behavior of quantum particles with the behavior of classical charged objects moving in Coulomb field. In classical physics, a negatively charged object accelerates due to the attraction of a positively charged object, and its kinetic energy gains with the decreasing distance from the positively charged object. However, in quantum physical model of the Hydrogen atom with an electron interacting with nucleus by Coulomb force, for a given eigenstate of electron with fixed energy, different spatial values (the distance from the nucleus) of electron correspond to the same energy. Similarly, Eq. (4.10) represents the eigenstates (or eigen-functions) of Eq. (2.11) with given quantum numbers ( $\omega,l,m$ ) and these quantum numbers do not change during the wave propagation measured in the same coordinate system.

For the second simpler case with $v_B \neq 0$, suppose that the gravitational field is weak and we can take the flat-limit approximation $R_\omega(r_B) \approx e^{-i\omega r_B}$ , then apply a trial solution with $\beta_{\omega\omega'} = 0$ and $\alpha_{\omega\omega'}$ is non-zero only at a certain $\omega'$ to be fixed, after substituting $R_\omega(r_B) \approx e^{-i\omega r_B}$ into Eq. (4.14) and Eq. (4.16) we get

$$\omega' = \omega[1 + v_B(t)]\frac{dt}{dt'} \tag{4.20}$$

in which $\frac{dt}{dt'}$ is given by Eq. (4.9). Note that Eq. (4.20) is valid under the condition that the strength of the gravitational field is weak. For a strong gravitational field (such as near the black-holes), Bob will measure a range of spectrum given by Eq. (4.14) instead of a single frequency given by Eq. (4.20). Assuming Alice carries an atom and Bob carries an identical atom. Alice's atom emits a wave $R_\omega(r)e^{-i\omega t}$ which has the frequency $\omega$ measured in Alice's coordinate $(t,r)$ to Bob. And the frequency of the wave measured by Bob is $\omega'$ expressed as Eq. (4.20). According to MH, we know that the energy levels of Bob's atom measured in coordinate $(t',r')$ are exactly the same as Alice's atom measured in coordinate $(t,r)$, where the two coordinates are related by Eq. (4.8). Therefore, Bob's atom will emit or absorb a wave with the same frequency $\omega$ measured in $(t',r')$ coordinate. In order to make Bob's atom absorb this wave sent by Alice, we require $\omega' = \omega$ in Eq. (4.20), then the velocity $v_B$ can be obtained as

$$v_B^{\pm} = \frac{-q \pm \sqrt{q(q^2+q-1)}}{q+1} \tag{4.21}$$

in which $q \equiv 1 - \frac{2M}{r_B}$. Therefore, the velocity has two solutions which correspond to two absorption peaks.

## V. Time Dilation and Doppler Effect with Time-Dependent Velocity in FRW Space-Time

The FRW (Friedmann-Robertson-Walker) metric in two dimensions can be written as

$$g_{\mu\nu}(t,r) = \begin{pmatrix} -1 & 0 \\ 0 & \dfrac{e^{2Ht}}{1-\kappa r^2} \end{pmatrix} \tag{5.1}$$

in which $H$ represents the Hubble constant measured in coordinate $(t,r)$ and $\kappa$ can take values as $\kappa = -1$ (open universe), $\kappa = 0$ (flat universe) or $\kappa = 1$ (closed universe), respectively. Similarly, in coordinate $(t',r')$ we have the metric $g_{\mu\nu}(t',r')$ written in the same form as

$$g_{\mu\nu}(t',r') = \begin{pmatrix} -1 & 0 \\ 0 & \dfrac{e^{2H't'}}{1-\kappa r'^2} \end{pmatrix} \tag{5.2}$$

in which $H'$ represents the Hubble constant measured in coordinate $(t',r')$. For the time being, we do not assume $H = H'$ for any coordinate system $(t',r')$. According to Eq. (2.13) the metric transformations between the two coordinates can be given as

$$g_{00} = -1 = -(\frac{\partial t'}{\partial t})^2 + \frac{e^{2H't'}}{1-\kappa r'^2}(\frac{\partial r'}{\partial t})^2, \tag{5.3a}$$

$$g_{01} = g_{10} = 0 = -\frac{\partial t'}{\partial t}\frac{\partial t'}{\partial r} + \frac{e^{2H't'}}{1-\kappa r'^2}\frac{\partial r'}{\partial t}\frac{\partial r'}{\partial r}, \tag{5.3b}$$

$$g_{11} = \frac{e^{2Ht}}{1-\kappa r^2} = -(\frac{\partial t'}{\partial r})^2 + \frac{e^{2H't'}}{1-\kappa r'^2}(\frac{\partial r'}{\partial r})^2 \tag{5.3c}$$

and Eq. (5.3) has an IPI solution which is given in Appendix C.

Now we consider two observers in flat universe ( $\kappa = 0$ ), Alice who carries the atomic clock A sets the coordinate $(t, r)$, and Bob who carries an identical clock B sets the coordinate $(t', r')$. At time $t = 0$ and $t' = 0$, Bob starts to move with velocity $v_B(t) \equiv \frac{dr_B}{dt}$. Next we can apply CTMLM, that is, the local metric value for the clock B measured by Bob equals to the local metric value for clock A measured by Alice, then Bob's coordinate system $(t', r')$ needs to satisfy

$$g'_{\mu\nu}(t', r'_B) = g_{\mu\nu}(t, r_A) \tag{5.4}$$

in which $g'_{\mu\nu}(t', r'_B)$ measured by Bob is the local metric where clock B locates and $g_{\mu\nu}(t, r_A)$ measured by Alice is the local metric where clock A locates. Again, for local experiments performed at $r'_B$, Bob does not need to specify metric values at other locations, the relationship of Eq. (5.4) is all we need to know. For convenience in the following discussions, assuming that Bob sets a coordinate system given by Eq. (5.2) which extends throughout the entire space-time, note that in this scenario, $(t', r')$ will be IPD functions of $(t, r)$ for non-zero velocity $v_B(t)$. Therefore, we have $e^{2Ht'} = e^{2Ht}$ and for the ticking events of clock B in flat universe, Eq. (5.3) become

$$-1 = -(\frac{\partial t'}{\partial t})^2 + e^{2Ht'}(\frac{\partial r'}{\partial t})^2, \tag{5.5a}$$

$$0 = -\frac{\partial t'}{\partial t}\frac{\partial t'}{\partial r} + e^{2Ht'}\frac{\partial r'}{\partial t}\frac{\partial r'}{\partial r}, \tag{5.5b}$$

$$e^{2Ht} = -(\frac{\partial t'}{\partial r})^2 + e^{2Ht'}(\frac{\partial r'}{\partial r})^2. \tag{5.5c}$$

For the first order differential relations, we have

$$dt' = \frac{\partial t'}{\partial t}dt + \frac{\partial t'}{\partial r}dr, \tag{5.6a}$$

$$dr' = \frac{\partial r'}{\partial t}dt + \frac{\partial r'}{\partial r}dr. \tag{5.6b}$$

If we treat the ticking of clock B as physical events, we have

$$dr' = (\frac{\partial r'}{\partial t} + v_B\frac{\partial r'}{\partial r})dt = 0 \Rightarrow \frac{\partial r'}{\partial t} + v_B\frac{\partial r'}{\partial r} = 0. \tag{5.7}$$

Combine Eq. (5.7) with Eq. (5.5), we can obtain the time dilation in expanding flat universe as (see Appendix C)

$$dt' = \sqrt{1 - e^{2Ht} v_B^2(t)} dt \tag{5.8}$$

in which $t$ is given by clock A and $t'$ is given by clock B.

For the mode solution of Eq. (2.11) with FRW metric in flat universe, we obtain the equation

$$3H \frac{\partial \psi}{\partial t} + \frac{\partial^2 \psi}{\partial t^2} = e^{-2Ht} \left( \frac{\partial^2 \psi}{\partial x^2} + \frac{\partial^2 \psi}{\partial y^2} + \frac{\partial^2 \psi}{\partial z^2} \right). \tag{5.9}$$

We can write the mode solution of Eq. (5.9) in $(t, \vec{x})$ coordinate as

$$\psi(t, \vec{x}) = \sum_{\vec{k}} [a_{\vec{k}} f_{\vec{k}}(t, \vec{x}) + a_{\vec{k}}^\dagger f_{\vec{k}}^*(t, \vec{x})]. \tag{5.10}$$

Then the mode is given by $f_{\vec{k}}(t, \vec{x}) = (2Ve^{3Ht})^{-1/2} e^{i\vec{k}\cdot\vec{x}} h_{\vec{k}}(t)$ in which $Ve^{3Ht}$ can be regarded as the physical volume of the cube on which the periodic boundary condition is imposed. We make the adiabatic approximation for $h_{\vec{k}}(t)$ as [29,32]

$$h_{\vec{k}}(t) = \omega_k(t)^{-1/2} \exp[-i \int^t \omega_k(t'') dt''] \tag{5.11}$$

in which the frequency is given by $\omega_k(t) \equiv |\vec{k}| e^{-Ht}$. Then $f_{\vec{k}}(t, \vec{x})$ in Eq. (5.10) can be given by

$$f_{\vec{k}}(t, \vec{x}) = (2V |\vec{k}|)^{-1/2} \exp[i\vec{k} \cdot \vec{x} + i \frac{\omega_k(t)}{H} - Ht]. \tag{5.12}$$

Similarly, in Bob's coordinate $(t', \vec{x}')$ with the metric given by Eq. (5.2), we get

$$\psi(t', \vec{x}') = \sum_{\vec{k}'} [a_{\vec{k}'} f_{\vec{k}'}(t', \vec{x}') + a_{\vec{k}'}^\dagger f_{\vec{k}'}^*(t', \vec{x}')]. \tag{5.13}$$

The wave $f_{\vec{k}}(t, \vec{x})$ can be decomposed into modes with positive and negative frequencies in $(t', \vec{x}')$ coordinate as

$$f_{\vec{k}}(t, \vec{x}) = \sum_{\vec{k}'} [\alpha_{\vec{k}\vec{k}'} f_{\vec{k}'}(t', \vec{x}') + \beta_{\vec{k}\vec{k}'} f_{\vec{k}'}^*(t', \vec{x}')]. \tag{5.14}$$

Now suppose Bob is moving in $x$ direction and Alice sends a wave $f_{\vec{k}}(t, \vec{x})$ to Bob with $k_y = k_z = 0$, when the wave reaches Bob at the location $x_B(t)$ measured by Alice, we have

$$f_k[t, x_B(t)] = \sum_{k'} [\alpha_{kk'} f_{k'}(t', x'_B) + \beta_{kk'} f_{k'}^*(t', x'_B)] \tag{5.15}$$

in which $x'_B$ is the spatial value of Bob measured in coordinate $(t', \vec{x}')$. After an infinitesimal time interval $dt$, we get

$$f_k[t+dt, x_B(t)+v_B(t)dt] = \sum_{k'}[\alpha_{kk'} f_{k'}(t'+dt', x'_B) + \beta_{kk'} f^*_{k'}(t'+dt', x'_B)]. \qquad (5.16)$$

Expanding Eq. (5.16) to first order, we obtain

$$f_k[t, x_B(t)] + [\frac{\partial f_k}{\partial t} + \frac{\partial f_k}{\partial x} v_B(t)]dt = \sum_{k'}[\alpha_{kk'}(f_{k'} + \frac{\partial f_{k'}}{\partial t'}dt') + \beta_{kk'}(f^*_{k'} + \frac{\partial f^*_{k'}}{\partial t'}dt')] \qquad (5.17)$$

in which $\frac{\partial f_k}{\partial t}$ can be obtained from Eq. (5.12) as

$$\frac{\partial f_k}{\partial t} = -(2Vk)^{-1/2}(ike^{-Ht} + H)\exp[ikx + i\frac{\omega_k(t)}{H} - Ht]. \qquad (5.18)$$

For large wave vector $k$ and small $H$, we make the approximation as

$$\frac{\partial f_k}{\partial t} \approx -ike^{-Ht}(2Vk)^{-1/2}\exp[ikx + i\frac{\omega_k(t)}{H} - Ht] = -ike^{-Ht} f_k(t,x). \qquad (5.19)$$

We apply a trial solution with $\beta_{kk'} = 0$ and $\alpha_{kk'} \neq 0$ only at one certain $k'$ which is to be fixed, thus combing Eq. (5.15) with Eq. (5.17) we have

$$f_k(t, x_B) = \alpha_{kk'} f_{k'}(t', x'_B), \qquad (5.20a)$$

$$f_k(t, x_B) + (v_B - e^{-Ht})ikf_k dt = \alpha_{kk'}(1 - i\omega' dt') f_{k'}(t', x'_B). \qquad (5.20b)$$

We get $\omega' = \omega(t)(1 - v_B e^{Ht})\frac{dt}{dt'}$ in which $\frac{dt}{dt'}$ is given by Eq. (5.8), at last, the red-shift of the wave is

$$\omega' = \omega(t)\sqrt{\frac{1 - v_B(t)e^{Ht}}{1 + v_B(t)e^{Ht}}}. \qquad (5.21)$$

Now we can see that the red-shift of the wave measured by Bob is caused by two factors, the first one is the cosmic red-shift expressed as $\omega_k(t) \equiv ke^{-Ht}$ which decreases with the expansion of the metric, and the second one is caused by the Doppler-shift shown in the square root. Again, Eq. (5.21) is valid only for large frequencies in the case of a non-zero velocity $v_B(t)$ because of the approximation we make in Eq. (5.19), for small frequencies Bob will measure a range of spectrum given by Eq. (5.15). In the case of $v_B(t) = 0$, one can verify that Eq. (5.21) holds even for small wave vector $k$. That is, we do not have to make the approximation Eq. (5.19) due to $\frac{dt}{dt'} = 1$ for two relatively static clocks.

At time $t = 0$, suppose that an atom carried by Alice emits a wave $f_k(t,x)$ to Bob, Bob who carries an identical atom is rest at location $x_B$ measured by Alice, and Bob receives this wave at time $t=t_1$ (this $t_1$ is measured by clock A). For the time-dependent metric, we cannot provide the argument from Eq. (2.17) to Eq. (2.27) to show that the mode frequency $\omega$ (measured by Alice) emitted by atoms at earlier time equals to the mode frequency $\omega'$ (measured by Bob) emitted by atoms at a later time. Rather, we make an assumption that the photon emitted by atomic transitions at different times has the same initial frequency, that is, the expanding universe does not change the energy levels of atoms, and some arguments supporting this assumption can be seen in other works [33,34]. Therefore, Bob's atom will emit or absorb a wave with the same initial frequency $\omega$ emitted by Alice at $t=0$. Thus, in order to make Bob's atom absorb this wave sent by Alice, Bob needs to move with a velocity $v_B(t_1)$ at time $t=t_1$ and we require $\omega'=\omega$ in Eq. (5.21). Therefore, we get

$$e^{-Ht_1}\sqrt{\frac{1-v_B(t_1)e^{Ht_1}}{1+v_B(t_1)e^{Ht_1}}}=1\,. \tag{5.22}$$

Then velocity $v_B(t_1)$ can be obtained as

$$v_B(t_1)=\frac{e^{-2Ht_1}-1}{e^{-Ht_1}+e^{Ht_1}}\,. \tag{5.23}$$

For the wave $f_k(t,x)$ traveling in null trajectory, that is $ds^2=-dt^2+e^{2Ht}dx^2=0$, we can calculate the time $t_1$ by the relation

$$\int_0^{t_1}\exp(-Ht)dt=\int_0^{x_B}dx\,. \tag{5.24}$$

Thus we get

$$t_1=\frac{1}{H}\ln\frac{1}{1-Hx_B}\,. \tag{5.25}$$

Therefore, if we require Bob's atom, which is at a spatial distance $x=x_B$ measured by Alice, to absorb this wave, he needs to move with a velocity $v_B(t_1)$ given by Eq. (5.23) at time $t_1$ and this $t_1$ is measured by clock A (if Bob is at rest before time $t_1$, then this $t_1$ is also given by clock B).

## VI. Conclusions

As a conclusion of this paper, in section II, we introduced a general coordinate transformation in curved space-times which is an analogy for the Lorentz transformation in Minkowski space-time, and such coordinate transformation is named as CTMPC. Moreover, we demonstrated that a coordinate transformation maintaining the local metric, which is named as CTMLM, is one type of CTMPC for atomic clocks. Meanwhile, the mathematical forms of the formulas and equations of QFT will be unchanged for CTMLM, this observation is crucial in studying the time dilation and Doppler effect with a time-dependent relative velocity in curved space-time. In sections III, IV and V, we studied CTMLM in Minkowski, Schwarzschild and FRW space-times, respectively. We showed that in order to remain the local metric value, the transformed coordinate parameters $(t',x')$ should become IPD functions of $(t,x)$. Meanwhile, the time dilation with an arbitrary time-dependent relative velocity is obtained in these space-times and the results agree with the clock hypothesis. For Minkowski space-time, we showed that if Alice sends a single-frequency wave to Bob who is moving with an arbitrary time-dependent velocity, Bob can always measure a single-frequency wave with a Doppler-shifted frequency value. However, in curved space-times such as Schwarzschild and FRW, if Alice sends a single-frequency wave to Bob, Bob may measure a broadened spectrum except under some certain conditions, Bob can still measure a single-frequency wave under these conditions. It is worth noting that the range of spectrum measured by Alice does not change, only the frequency range perceived by Bob's atoms as receivers of the waves may be broadened. And this spectra line broadening effect is not in the same category as other effects caused by thermal motions or pressures [35,36]. Instead, it is an intrinsic feature caused by the curvature of space-time itself, and this feature needs to be taken into consideration for future astrophysical observations.

**Acknowledgement:**

This work is supported by National Science Foundation of China under Grant No. 11847307.

APPENDIX A

The velocity $v_B(t)$ of Bob can be expressed as

$$v_B(t)=\begin{pmatrix} a_B t, \quad 0\le t<t_1 \\ 2a_B t_1 - a_B t, \quad t_1\le t<3t_1 \\ a_B t - 4a_B t_1, \ 3t_1 \le t < 4t_1 \end{pmatrix}. \tag{A1}$$

Since $dx'=0$ along path II, we get $x'_{II}(\tilde{x})\equiv\int_{path\,II}dx'=0$. From Eq. (3.2a) we have $dt'=[\cosh f(\tilde{x})-v_B\sinh f(\tilde{x})]dt$. Since $v_B=\tanh f(\tilde{x})$, then $\cosh f(\tilde{x})=\frac{1}{\sqrt{1-v_B^2}}$ and $\sinh f(\tilde{x})=\frac{v_B}{\sqrt{1-v_B^2}}$, we obtain

$$dt'=\sqrt{1-v_B^2}\,dt. \tag{A2}$$

Note that here $v_B$ is time-dependent as shown in Eq. (A1). Now by integrating Eq. (A2) along path II, we get

$$t'_{II}(\tilde{x}) = \int_0^{4t_1} \sqrt{1-v_B^2}\,dt = 4\int_0^{t_1} \sqrt{1-a_B^2 t^2}\,dt = \frac{2}{a_B}\arcsin a_B t_1 + 2t_1\sqrt{1-a_B^2 t_1^2}\,. \tag{A3}$$

For the integration of path I, we have $dx = 0$, therefore from Eq. (3.11) we obtain

$$t'_I(\tilde{x}) = \int_0^{4t_1} \cosh f(\tilde{x})dt = 4\int_0^{t_1} \frac{1}{\sqrt{1-v_B^2}}dt = \frac{4}{a_B}\arcsin a_B t_1, \tag{A4}$$

$$x'_I(\tilde{x}) = \int_0^{4t_1} -\sinh f(\tilde{x})dt = \int_0^{4t_1} \frac{-v_B}{\sqrt{1-v_B^2}}dt = 0. \tag{A5}$$

## APPENDIX B

If we treat the ticking of clock B (outside of event horizon) as the physical event, since these events occur along Bob's co-moving frame $(t', r')$, we get

$$dr' = (\frac{\partial r'}{\partial t} + v_B\frac{\partial r'}{\partial r})dt = 0 \Rightarrow \frac{\partial r'}{\partial t} + v_B\frac{\partial r'}{\partial r} = 0\,. \tag{B1}$$

To simplify the mathematical expressions, we define $q \equiv 1 - \frac{2M}{r}$, from Eq. (4.8) we obtain

$$\frac{\partial r'}{\partial t} = \pm\sqrt{(\frac{\partial t'}{\partial t})^2 - q}\,, \quad \frac{\partial t'}{\partial r} = \pm\sqrt{(\frac{\partial t'}{\partial t})^2 / q^2 - \frac{1}{q}}\,, \quad \frac{\partial r'}{\partial r} = \frac{\partial t'}{\partial t}\frac{1}{q}\,. \tag{B2}$$

Note that we will take positive values for $\frac{\partial r'}{\partial r}$ and $\frac{\partial t'}{\partial t}$ (Bob arranges his coordinate in the same direction as Alice), the sign of $\frac{\partial r'}{\partial t}$ and $\frac{\partial t'}{\partial r}$ will depend on the sign of $v_B(t)$. For $v_B(t) > 0$, according to Eq. (B1) and Eq. (B2), at time $t$ we obtain

$$\frac{\partial t'}{\partial t} = \sqrt{\frac{q^3}{q^2 - v_B^2(t)}}\,. \tag{B3}$$

We can give the time relationship between two clocks as

$$dt' = (\frac{\partial t'}{\partial t} + v_B \frac{\partial t'}{\partial r})dt = \sqrt{\frac{q^2 - v_B^2(t)}{q}} dt . \tag{B4}$$

In the case that the clock B is at rest, i.e. $v_B(t) = 0$, we get the time dilation due to gravity as

$$dt' = \sqrt{1 - \frac{2M}{r}} dt . \tag{B5}$$

Therefore, the atomic clock near the gravity source runs slower than a clock located far away from the gravity field.

## APPENDIX C

To simplify the expressions, we define $p \equiv \frac{e^{2Ht}}{1-\kappa r^2}$ and $p' \equiv \frac{e^{2H't'}}{1-\kappa r'^2}$, from Eq. (5.3) we have

$$\frac{\partial r'}{\partial t} = \pm\sqrt{\frac{(\frac{\partial t'}{\partial t})^2 - 1}{p'}}, \quad \frac{\partial t'}{\partial r} = \pm\sqrt{p(\frac{\partial t'}{\partial t})^2 - p}, \quad \frac{\partial r'}{\partial r} = \pm\frac{\partial t'}{\partial t}\sqrt{\frac{p}{p'}} . \tag{C1}$$

Next we apply a trial solution $t' = t + \varepsilon$ in which $\varepsilon$ is a constant and $H' = H$, we obtain $\frac{\partial t'}{\partial t} = 1$, $\frac{\partial r'}{\partial t} = \frac{\partial t'}{\partial r} = 0$ and $\frac{\partial r'}{\partial r} = e^{-H\varepsilon}\sqrt{\frac{1-\kappa r'^2}{1-\kappa r^2}}$ (here we take the positive root). Now we can see that if $r'$ is not a function of $t$, then Eq. (4.5) are satisfied. We integrate parameter $r$ to get the relation between $r'$ with $r$ for different $\kappa$ as

$$\arcsin r' = e^{-H\varepsilon}(\arcsin r + a), \qquad \kappa = -1 \tag{C2}$$

$$r' = e^{-H\varepsilon}(r + a), \qquad \kappa = 0 \tag{C3}$$

$$\ln(r' + \sqrt{r'^2 + 1}) = e^{-H\varepsilon}[\ln(r + \sqrt{r^2 + 1}) + a], \qquad \kappa = 1 \tag{C4}$$

in which $a$ is an integration constant. Therefore, Eq. (C2) to Eq. (C4) together with $t' = t + \varepsilon$ is an IPI coordinate transformation from $(t, r)$ to $(t', r')$ and the ticking rate for two atomic clocks rest at two different positions is the same. In particular for flat universe ($\kappa = 0$) with $\varepsilon = 0$, we get a translation symmetry as

$$r' = r + a , \qquad t' = t \tag{C5}$$

which is the same as spatial translation in Minkowski space-time.

For Bob moving with velocity $v_B(t) > 0$ in flat universe, Eq. (C1) together with Eq. (5.7) gives

$$v_B(t) = -\frac{\partial r'}{\partial t} / \frac{\partial r'}{\partial r} = \sqrt{\frac{(\frac{\partial t'}{\partial t})^2 - 1}{e^{2Ht}(\frac{\partial t'}{\partial t})^2}} \Rightarrow \frac{\partial t'}{\partial t} = \sqrt{\frac{1}{1-e^{2Ht} v_B^2(t)}} \,. \quad \text{(C6)}$$

Note that we take positive root for $\frac{\partial t'}{\partial t}$ and $\frac{\partial r'}{\partial r}$, that is, Bob sets the positive direction of coordinate $(t', r')$ the same as $(t, r)$. Therefore, substitute Eq. (C6) into Eq. (C1) and Eq. (5.6) we get

$$dt' = [\sqrt{\frac{1}{1-e^{2Ht} v_B^2(t)}} - e^{2Ht} v_B^2(t) \sqrt{\frac{1}{1-e^{2Ht} v_B^2(t)}}] dt = \sqrt{1-e^{2Ht} v_B^2(t)} dt \quad \text{(C7)}$$

in which $t$ is the time displayed by clock A and $t'$ is the time displayed by clock B. This result is the time dilation of moving atomic clocks in expanding flat universe.